\documentclass[conference]{IEEEtran}
\IEEEoverridecommandlockouts
\usepackage{cite}
\usepackage{amsmath,amssymb,amsfonts}
\usepackage{algorithmic}
\usepackage{graphicx}
\usepackage{textcomp}
\usepackage{xcolor}
\usepackage{booktabs}
\usepackage{listings}
\usepackage{subfig}
\usepackage{caption}
\usepackage[hyphens]{url}
\usepackage[colorlinks=true,citecolor=red]{hyperref}
\usepackage{todonotes}
\usepackage{float}
\usepackage{amsfonts}
\usepackage{svg}
\usepackage{tikz}
\usepackage{amssymb}% http://ctan.org/pkg/amssymb
\usepackage{pifont}% http://ctan.org/pkg/pifont
\usetikzlibrary{patterns}
\usetikzlibrary{shapes.geometric}
\usepackage[para]{threeparttable}
\usepackage[]{quoting}

\newcommand*\blackbox{
    \begin{tikzpicture}
    \node[circle,draw=black,fill=black,inner sep=0pt,minimum size=6pt] {};
    \end{tikzpicture}
}

\newcommand*\whitebox{
    \begin{tikzpicture}
    \node[circle,draw=black,fill=white,inner sep=0pt,minimum size=6pt] {};
    \end{tikzpicture}
}

\newcommand*\graybox{
    \begin{tikzpicture}
    \node[circle,draw=black,fill=white,pattern=crosshatch,inner sep=0pt,minimum size=6pt] {};
    \end{tikzpicture}
}

\newcommand*\modelextraction{
    \begin{tikzpicture}[square/.style={regular polygon,regular polygon sides=4}]
    \node[square,draw=black,fill=white,inner sep=0pt,minimum size=8pt] {};
    \end{tikzpicture}
}

\newcommand*\privacyleakage{
    \begin{tikzpicture}[square/.style={regular polygon,regular polygon sides=4}]
    \node[square,draw=black,fill=black,inner sep=0pt,minimum size=8pt] {};
    \end{tikzpicture}
}

\newcommand*\adversarial{
    \begin{tikzpicture}[square/.style={regular polygon,regular polygon sides=4}]
    \node[square,draw=black,fill=white,pattern=crosshatch,inner sep=0pt,minimum size=8pt] {};
    \end{tikzpicture}
}

\definecolor{dkgreen}{rgb}{0,0.6,0}
\definecolor{gray}{rgb}{0.5,0.5,0.5}
\definecolor{mauve}{rgb}{0.58,0,0.82}
\definecolor{codegreen}{rgb}{0,0.6,0}
\definecolor{codegray}{rgb}{0.5,0.5,0.5}
\definecolor{codepurple}{rgb}{0.58,0,0.82}
\definecolor{backcolour}{rgb}{0.95,0.95,0.92}

\lstdefinestyle{myStyle}{
    belowcaptionskip=1\baselineskip,
    frame=none,
    basicstyle=\footnotesize\ttfamily,
    backgroundcolor=\color{backcolour},   
    commentstyle=\color{codegreen},
    keywordstyle=\color{magenta},
    numberstyle=\tiny\color{codegray},
    stringstyle=\color{codepurple},
    breakatwhitespace=true,
    breaklines=true,                 
    keepspaces=true,                 
    numbers=left,       
    numbersep=2pt,
    showspaces=false,                
    showstringspaces=false,
    showtabs=false,                  
    tabsize=2,
}

\def\BibTeX{{\rm B\kern-.05em{\sc i\kern-.025em b}\kern-.08em
    T\kern-.1667em\lower.7ex\hbox{E}\kern-.125emX}}

\begin{document}

\title{On the Evaluation of User Privacy in Deep Neural Networks using Timing Side Channel \\
{\footnotesize \textsuperscript{*}Note: An extended version of this paper has been accepted to CHES 2023 \cite{DBLP:journals/tches/ShuklaABMM23}}
}

\author{\IEEEauthorblockN{Shubhi Shukla}
\IEEEauthorblockA{\textit{IIT Kharagpur}\\
Kharagpur, India \\}
\and
\IEEEauthorblockN{Manaar Alam}
\IEEEauthorblockA{\textit{NYU Abu Dhabi}\\
Abu Dhabi, UAE \\}
\and
\IEEEauthorblockN{Sarani Bhattacharya}
\IEEEauthorblockA{\textit{IMEC}\\
Leuven, Belgium \\}
\and
\IEEEauthorblockN{Debdeep Mukhopadhyay}
\IEEEauthorblockA{\textit{IIT Kharagpur}\\
Kharagpur, India \\}
\and
\IEEEauthorblockN{Pabitra Mitra}
\IEEEauthorblockA{\textit{IIT Kharagpur}\\
Kharagpur, India \\}
}

\maketitle

\begin{abstract}

Recent Deep Learning (DL) advancements in solving complex real-world tasks have led to its widespread adoption in practical applications. However, this opportunity comes with significant underlying risks, as many of these models rely on privacy-sensitive data for training in a variety of applications, making them an overly-exposed threat surface for privacy violations. Furthermore, the widespread use of cloud-based \textit{Machine-Learning-as-a-Service (MLaaS)} for its robust infrastructure support has broadened the threat surface to include a variety of remote \textit{side-channel attacks}.
In this paper, we first identify and report a novel data-dependent timing side-channel leakage (termed \textit{Class Leakage}) in DL implementations originating from \textit{non-constant time} branching operation in a widely used DL framework, \textit{PyTorch}. We further demonstrate a practical inference-time attack where an adversary with \textit{user privilege} and \textit{hard-label black-box access} to an MLaaS can exploit Class Leakage to compromise the privacy of MLaaS users. DL models are vulnerable to \textit{Membership Inference Attack} (MIA), where an adversary's objective is to deduce whether any particular data has been used while training the model. In this paper, as a separate case study, we demonstrate that a DL model secured with \textit{differential privacy} (a popular countermeasure against MIA) is still vulnerable to MIA against an adversary exploiting Class Leakage. We develop an easy-to-implement countermeasure by making a \textit{constant-time} branching operation that alleviates the Class Leakage and also aids in mitigating MIA. 
We have chosen two standard benchmarking image classification datasets, \textit{CIFAR-10} and \textit{CIFAR-100} to train five state-of-the-art pre-trained DL models, over two different computing environments having \textit{Intel Xeon} and \textit{Intel i7} processors to validate our approach.

\end{abstract}


\section{Introduction}
In recent years, we have seen an outburst of research using Deep Learning (DL) in both industry and academia because of its undeniable performances in long-standing AI tasks in various domains, such as image recognition~\cite{DBLP:conf/cvpr/HeZRS16}, machine translation~\cite{DBLP:journals/corr/BahdanauCB14}, malware detection~\cite{DBLP:journals/access/VinayakumarASPV19}, and autonomous driving~\cite{DBLP:journals/jfr/GrigorescuTCM20}. Crowd-sourcing technology giants like Google, Facebook, Amazon, and others collect massive amounts of training data from their users and deploy personalized DL applications on a large scale. While the utility of DL is unquestionable, the training data behind its success present serious privacy concerns. The massive database of images, audio, and video collected from millions of individuals is an abundant source of privacy-related risks. Moreover, DL is also used in many privacy-preserving domains, like medical data analysis~\cite{DBLP:journals/corr/abs-1711-05225}, where sharing data about individual or entity without suitable permission is not allowed by law or regulation.

The complexity, dynamism, and volume of data in the real world have recently promoted cloud-based Machine Learning as a Service (MLaaS)~\cite{DBLP:conf/icmla/RibeiroGC15}, which provides infrastructural support of powerful computing resources and substantial domain expertise to train efficient DL models. MLaaS operates on data obtained from the clients for building DL models, saving the burden of cognate cost and time to train these models from scratch and increasing their availability to larger audiences. Amazon Machine Learning, Microsoft Azure, Google Cloud Platform, and IBM Cloud are key players in the recent market providing MLaaS through prediction APIs. MLaaS is even getting popular among healthcare software solutions that operate on private medical datasets~\cite{rayome_2017,mejia_2019}. The growing popularity and easy availability of MLaaS have increased the concerns to protect user privacy even more.

There is a relevant field of study pertinent to privacy-preserving DL algorithms, where the primary assumption is that the private data is only known to individual users, not even by the entity that uses the data to train the model. These algorithms use computationally intensive secure multi-party computation~\cite{DBLP:journals/corr/abs-2109-00984,DBLP:conf/uss/KotiPPS21} and homomorphic encryption~\cite{DBLP:conf/icml/Gilad-BachrachD16,DBLP:conf/nips/Lou019} to algorithmically protect user data from entities other than the user itself. \textbf{However, in this paper, we try to address the privacy issues in an MLaaS framework from the perspective of a passive adversary.} We consider that a trusted entity trains a private dataset and designs an implementation for the trained model. We assume that a passive adversary has \textit{hard-label black-box API access}\footnote{In hard-label black-box access, a client can only obtain actual labels and does not get any knowledge of probabilities associated with predicted labels.} to the trained model and \textit{can obtain information by exploiting the weakness in the implementation of such models} (popularly known as side-channel information). We attempt to address the security issues in this scenario, which is practical and different from the context of the aforementioned field of study.

We have recently seen a considerable number of research that exploit side-channel information in remote environments to reverse-engineer architecture and parameters of Deep Neural Networks (DNN) that are commercialized and kept undisclosed. \textit{Hong~et~al.} used the cache-based side-channel during the inference phase to reconstruct the crucial architectural secret of the victim DNN~\cite{DBLP:journals/corr/abs-1810-03487}. \textit{Yan~et~al.} exploited the Generalized Matrix Multiply (GEMM) of the victim DNN implementation using cache-based side-channel to extract its crucial structural secret~\cite{DBLP:conf/uss/YanFT20}. \textit{Naghibijouybari~et~al.} exploited shared GPU resources coupled with hardware performance counters to extract secret internal structure of the victim DNN~\cite{DBLP:conf/ccs/Naghibijouybari18}. \textit{Wei~et~al.} exploited the GPU context-switch side-channel to steal the fine-grained architectural secret of the victim DNN~\cite{DBLP:conf/dsn/WeiZZLF20}. \textit{Rankin~et~al.} proposed a methodology to steal crucial parameters of victim DNN by exploiting rowhammer fault injection on DRAM modules~\cite{DBLP:journals/corr/abs-2111-04625}. \textit{Duddu~et~al.} exploited the timing side-channel to construct an optimal substitute architecture of the victim DNN~\cite{DBLP:journals/corr/abs-1812-11720}. \textbf{While these works predominantly focus on reverse-engineering commercialized DL models using side-channel leakages, in this paper, we attempt to exploit side-channel leakage to compromise user privacy.} In this context, we define a term \textit{Class-Leakage} as side-channel-based information leakage from \textit{DL implementations} that aids an adversary in distinguishing unknown labels of different inputs without explicitly accessing the model or the input data. Information on the input label is directly linked to the sensitive information of a user, highlighting a critical privacy concern in DL implementations.

The continuous rise of DL has propelled the growth of various open-source libraries, like Tensorflow~\cite{DBLP:conf/osdi/AbadiBCCDDDGIIK16}, Keras~\cite{chollet2015keras}, PyTorch~\cite{DBLP:conf/nips/PaszkeGMLBCKLGA19}, Theano~\cite{DBLP:journals/corr/Al-RfouAAa16}, etc., for efficient data flow while implementing DL models. \textbf{In this paper, we primarily focus on PyTorch and identified an implementation vulnerability typically responsible for Class-Leakage through timing side-channel\footnote{The simplicity, ease of use, dynamic computational graph, and efficient memory usage have recently made PyTorch one of the most sought-after libraries for several organizations to implement industrially standard DL applications~\cite{sameer_2020,scott_2020}. This paper also motivates that these applications are likewise exposed to the discussed vulnerability.}.} We demonstrate that operation of the \textit{Max Pooling} module in a Convolutional Neural Network (CNN) (a class of DNN popularly used for image classification tasks) using PyTorch is vulnerable to input-dependent timing side-channel leakage due to improper \textit{non-constant time} implementation of branching instruction\footnote{We reported the vulnerability to the Meta (Facebook) AI Research team (developer of the PyTorch library), who also acknowledged the vulnerability. We discuss the vulnerability disclosure in details later in Section~\ref{section:fb_ack}.}. Further, we also demonstrate that the vulnerability can be exploited during the inference phase of a DL model by an adversary having access to a \textit{manifest dataset} to compromise a user's privacy using a Multi-Layer Perceptron (MLP). The manifest dataset is a set of \textit{annotated data} that is apparent to the adversary, though not necessarily a subset of the original training dataset but resembles it sufficiently. While \textit{Duddu~et~al.}~\cite{DBLP:journals/corr/abs-1812-11720} exploited timing channel for reverse-engineering a commercialized DL model and \textit{Nakai~et~al.}~\cite{DBLP:journals/tches/NakaiSF21} exploited timing channel in an embedded platform for crafting \textit{adversarial examples}\footnote{A class of threats that adds visually imperceptible perturbations to the input data of a DNN to cause misclassification.}, \textbf{to the best of our knowledge, this is the first work to show that timing side-channel can also be used to compromise user privacy from a DL implementation in an MLaaS framework.} \textit{Alam~et~al.}~\cite{DBLP:conf/dac/AlamM19} and \textit{Wang~et~al.}~\cite{wang_2022} also demonstrated that DNNs are vulnerable to leaking label information of an input instance using cache-based side-channel. However, the method proposed by \textit{Alam~et~al.} requires \textit{super-user privilege} to access \textit{hardware performance counters} of the system executing DL implementations. On the other hand, the method proposed by \textit{Wang~et~al.} requires full access to the parameters of DL models (i.e., \textit{white-box} access) or needs to acquire the parameters using previous research on reverse engineering~\cite{DBLP:journals/corr/abs-1810-03487,DBLP:conf/uss/YanFT20,DBLP:conf/ccs/Naghibijouybari18,DBLP:conf/dsn/WeiZZLF20,DBLP:journals/corr/abs-2111-04625,DBLP:journals/corr/abs-1812-11720}. The super-user privilege of a system or the white-box access to DL models may not be practical in various applications where the security and privacy of users are of utmost importance. \textbf{In this paper, we assume that the adversary has hard-label black-box access to DL models from the user-space and can compromise privacy even without reverse-engineering any DNN parameters.} We summarise the contribution of this paper in comparison to the related works on remote\footnote{Several recent works also exploit side-channel information like power~\cite{DBLP:conf/acsac/WeiLLL018}, electromagnetic emanation~\cite{DBLP:conf/uss/BatinaBJP19,DBLP:conf/host/YuMYZJ20}, and off-chip memory access~\cite{DBLP:conf/dac/HuaZS18} to reverse engineer architectural secrets of DNNs, which require physical access to the model. We have considered a cloud-based remote MLaaS scenario in this paper and do not provide details of such works.} side-channel-based attacks on DL in Table~\ref{table:sota_comparison}.

\begin{table}[!t]
\resizebox{\linewidth}{!}{
\begin{threeparttable}
\centering
\caption{Summary of contribution of this work in comparison to related works on remote side-channel-based attacks on DL\label{table:sota_comparison}}
\begin{tabular}{|c|c|c|c|}
\hline
\textbf{Paper} & \textbf{Side-Channel} & \textbf{\begin{tabular}[c]{@{}c@{}}Adversary\\ Capability\end{tabular}} & \textbf{Objective} \\ \hline
\cite{DBLP:journals/corr/abs-1810-03487},\cite{DBLP:conf/uss/YanFT20},\cite{DBLP:conf/dac/AlamM19}\tnote{$\mathsection$},\cite{wang_2022} & Cache & \blackbox,\blackbox,\blackbox,\whitebox & \modelextraction,\modelextraction,\privacyleakage,\privacyleakage \\ \hline
\cite{DBLP:conf/ccs/Naghibijouybari18},\cite{DBLP:conf/dsn/WeiZZLF20} & GPU & \blackbox,\blackbox & \modelextraction,\modelextraction \\ \hline
\cite{DBLP:journals/corr/abs-2111-04625} & Memory & \graybox & \modelextraction \\ \hline
\cite{DBLP:journals/corr/abs-1812-11720},\cite{DBLP:journals/tches/NakaiSF21}\tnote{$\ddagger$}, \textbf{This Work} & Time & \blackbox,\blackbox,\blackbox & \modelextraction,\adversarial,\privacyleakage \\ \hline
\end{tabular}
\begin{tablenotes}
    \item [$\mathsection$] Requires super-user privilege.
    \item [$\ddagger$] Requires physical access.
\end{tablenotes}
\begin{tablenotes}
 \whitebox: Full knowledge of the DNN.
 \graybox: Only model architecture is known.\\
 \blackbox: Black-box access to the DNN.
\end{tablenotes}
\begin{tablenotes}
 \modelextraction: Model reverse-engineering.
 \adversarial: Craft adversarial examples.\\
 \privacyleakage: Leak label information.
\end{tablenotes}
\end{threeparttable}
}\vspace{-0.3cm}
\end{table}

Aside from the aforementioned threat to user privacy, we also consider an additional case study pertinent to \textit{Membership Inference Attack} (MIA)~\cite{DBLP:conf/sp/ShokriSSS17,DBLP:conf/icml/Choquette-ChooT21,DBLP:journals/tsc/TruexLGYW21}, where the objective of an adversary is to deduce whether an unknown data has been part of the dataset used while training a DL model. MIA can be staged without requiring any access to DNN parameters and simply by observing its output, causing critical privacy ramifications to users whose data records have been used to train the DL model. We have seen a significant number of efforts in recent literature to prevent MIA -- differential privacy has gained popularity among them because of its simplicity and competence in preventing MIA~\cite{DBLP:conf/ccs/AbadiCGMMT016,DBLP:journals/corr/abs-2007-11524,DBLP:conf/aaai/PapernotT0CE21}. \textbf{In this paper, we show that a DL model secured using differential privacy is still vulnerable against MIA if the adversary has additional information through timing side-channel leakage due to the improper implementation, as mentioned before.} We used the open-source library \textit{Opacus}~\cite{DBLP:journals/corr/abs-2109-12298} to train DL models with differential privacy using PyTorch and demonstrate its vulnerability against MIA using the timing side-channel-based Class-Leakage.

\textbf{In this work, we also thrive on implementing a simple countermeasure by making minimal changes to the existing PyTorch library to mitigate the correlation of timing side-channel with input data without affecting the accuracy of DL models.} With the addition of \textit{only two extra lines of codes} to the existing PyTorch library, we show that one can implement the Max Pooling operation in \textit{constant time} without affecting the accuracy of DL models. We demonstrate that the proposed modified constant-time Max Pooling module can successfully mitigate the input-dependent timing side-channel leakage existing in the current PyTorch-based implementation. We also demonstrate that the proposed constant-time implementation of the Max Pooling module successfully alleviates the vulnerability of differentially private DL models based on PyTorch (like Opacus) against MIA even if the adversary has additional information through timing side-channel.

\textbf{Contributions:} Our primary contributions through this paper are discussed as follows:
\begin{itemize}
    \item We identified an \textit{implementation vulnerability in the Max Pooling operation} of a CNN implemented using the \textit{PyTorch} library. The vulnerability results in \textit{data-dependent timing side-channel leakage} due to \textit{improper non-constant time implementation} of a branching instruction. The timing side-channel leads to \textit{Class-Leakage} that aids an adversary having only \textit{hard-label black-box access} and \textit{user-level privilege} in distinguishing unknown labels of different inputs without explicitly accessing the DL model or the input data.
    
    \item We demonstrate a methodology where an adversary with access to the \textit{manifest dataset} can exploit the \textit{Class-Leakage} vulnerability during the \textit{inference phase} of a DL model to \textit{compromise users' privacy by predicting unknown labels of the users' inputs} that are directly linked to their sensitive information using an MLP.
    
    \item We demonstrate that a DL model secured using differential privacy can \textit{still be vulnerable against MIA if the adversary has additional information through timing side-channel-based Class-Leakage}.
    
    \item We propose an \textit{easy-to-implement countermeasure} to develop a \textit{constant-time Max Pooling operation} by making minimal changes (\textit{only two extra lines of codes}) to the existing PyTorch library that does not affect the accuracy of DL models. We show that the countermeasure \textit{mitigates the Class-Leakage} and \textit{alleviates the timing side-channel-based vulnerability of differential-private DL models against MIA.}
    
    \item We evaluated all the experiments on standard image classification benchmarking datasets like \textit{CIFAR-10}~\cite{cifar10} and \textit{CIFAR-100}~\cite{cifar100} using state-of-the-art CNN models like \textit{AlexNet}~\cite{DBLP:conf/nips/KrizhevskySH12}, \textit{DenseNet121}~\cite{DBLP:conf/cvpr/HuangLMW17}, \textit{SqueezeNet}~\cite{DBLP:journals/corr/IandolaMAHDK16}, \textit{ResNet50}~\cite{DBLP:conf/cvpr/HeZRS16}, and \textit{VGG19}~\cite{DBLP:journals/corr/SimonyanZ14a}. In order to validate the generalizability of the method in different computing environments, we performed all the experiments both on Intel Xeon and Intel i7 processors.
\end{itemize}

The rest of the paper is organised as follows: 
Section~\ref{section:prelim} presents a brief overview of the necessary background required for understanding of this paper. Section~\ref{section:vulnerability} introduces the data-dependent timing side-channel leakage identified in PyTorch. Section~\ref{section:MLP_Attack} demonstrates a practical inference-time attack using the timing side-channel leakage. Section~\ref{section:DP_CL_CNN} illustrates the vulnerability of differential-private models against Membership Inference Attacks using the timing side-channel leakage. Section~\ref{section:countermeasure} discusses a proposed countermeasure to alleviate the timing side-channel leakage. Section~\ref{section:fb_ack} presents a brief discussion on disclosure of the vulnerability to Meta (Facebook) AI research. Finally, Section~\ref{section:conclusion} concludes the paper with a short discussion on practical impacts of this vulnerability.

\section{Preliminaries}
\label{section:prelim}
\subsection{Multilayer Perceptron}
Artificial Neural Networks (ANNs)~\cite{DBLP:series/ascas/Graupe13} are complex networks which comprise of neurons connected to each other. Each neural network comprises of an input layer, output layer and possibly some hidden layers as well if required. The simplest form of ANNs are feed-forward neural networks~\cite{DBLP:conf/colt/EldanS16}, in which there are no cycles in among the path of neurons, unlike some other networks like recurrent neural networks. Additionally, the data from neurons only moves ahead in one direction, from the input layer to the output layer. Multilayer perceptron (MLP) also comes under the category of feed forward neural networks. In general, a MLP has at least one hidden layer. The input to a MLP passes through each layer by doing numerous weighted matrix multiplications and then are passed through an activation function. Some common activation functions are sigmoid, tanh and ReLU. After these computations in each layer neuron of a layer, the outputs are forwarded as inputs to the next layer, and the process keeps repeating until the output layer is reached.
% An example of a four layer MLP is shown in Figure \ref{fig:mlp}.

% \begin{figure}[!ht]
% % \centerline{\includegraphics[width=0.4\textwidth]{CIFAR10_100_DP_red.png}}
% \centerline{\includesvg[width=0.3\textwidth]{mlp.svg}}
% \caption{An example of a multilayer perceptron with two hidden layers}
% \label{fig:mlp}
% \end{figure}

\subsection{Convolution Neural Networks}
Convolutional Neural Network (CNNs) \cite{CNN} are type of neural network specializing in processing Multidimensional matrix data like time-series data and image data. CNNs derive their unique advantage by their method of processing the data. They use a grid like structure to move around the data and extract useful features.  A critical advantage of using CNNs over fully connected DNNs is lower requirement for processing the data. A basic CNN architecture has three main layers: convolution, pooling and fully-connected layers.

% \begin{figure}[!ht]
% % \centerline{\includegraphics[width=0.4\textwidth]{CIFAR10_100_DP_red.png}}
% \centerline{\includesvg[width=0.45\textwidth]{ConvNet.svg}}
% \caption{An example of a Convolution Neural Network with Max-poo, convolution and fully-connected layers}
% \label{fig:conv}
% \end{figure}

\subsection{Max Pooling}
\begin{figure}[!t]
\centerline{\includegraphics[width=0.7\linewidth]{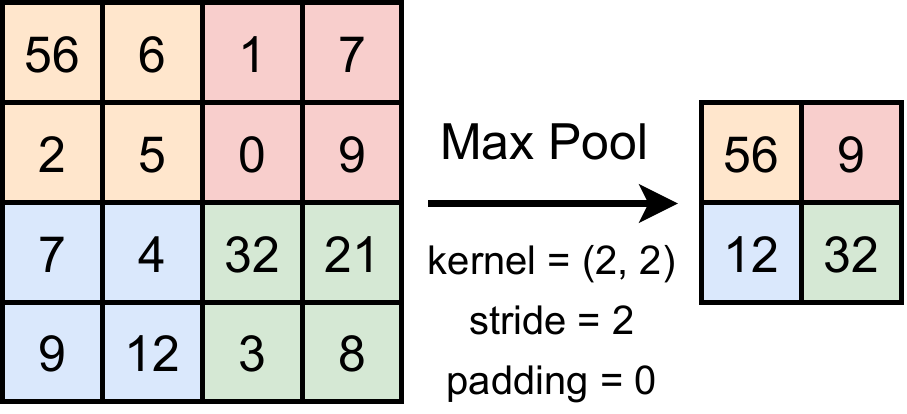}}
\caption[Max pooling operation example]{An example of max pooling operation\vspace{-0.4cm}}
\label{maxpool}
\end{figure}

Pooling layers \cite{DBLP:journals/corr/abs-2009-07485} in a CNN are used to reduce the size of the feature maps for further processing by consolidating the outputs of the previous layers. In general, there are two types of pooling functions; Max pool and Average pool. Pooling function is usually defined by three parameters: Kernel size, stride and padding. \emph{Kernel size} is the window size of the sub-matrix on which pooling operation will be applied. In case of max pooling, the maximum value among all elements in the window will be the output for that window. The \emph{stride} parameter defines the step size of the window both in the direction of rows and columns. \emph{Padding} parameter is generally used when we want to regulate the size of the output matrix. Padding adds additional rows and columns with zero values at the front, back top and bottom of the matrix. The width of the padding is determined by the padding parameter. Fig.~\ref{maxpool} shows an example of max pooling operation with stride $2$ and kernel size of $2 \times 2$.

\subsection{Timing Side-channel Analysis}
Attacks launched using timing side-channel leakages are one of the most common and simple side-channel attacks \cite{DBLP:conf/sp/HundWH13}. We say that there is timing leakage in a certain algorithm or process, if by observing the timing traces of that process the adversary is able to gain some secret information, which it should not have access to in an ideal case. The cause of timing leakages can be difference of execution times of different operations or instructions which may affect the overall execution time of the process in different settings. To illustrate the concept of timing side-channel leakage we take an example. Let's take a function $\mathcal{X}$, which takes one input and performs some operation on it. Let $p$ and $q$ be the only two possible inputs to function $\mathcal{X}$. Let $\mathcal{P}$ and $\mathcal{Q}$ be distributions of execution times of $\mathcal{X}$ with inputs $p$ and $q$ respectively. In an ideal scenario, the function should take same amount of time to execute with both inputs, hence $\mathcal{P}$ and $\mathcal{Q}$ should be statistically similar to each other. In case they are not similar we say that there is a timing leakage, since we can infer the input to the function $\mathcal{X}$ by getting the timing traces.

% \begin{figure}[!t]
% % \centerline{\includesvg[width=0.75\textwidth]{TimingLeakage.drawio.svg}}
% \centerline{\includegraphics[width=0.75\linewidth]{TimingLeakage.drawio.pdf}}
% % \centerline{\includesvg[width=4cm]{TimingLeakage.drawio.svg}}
% \caption{Timing Side-channel Leakage}
% \label{TimingLeakage}
% \end{figure}

\subsection{Differential Privacy}
In today's time government organizations, technology industry, healthcare industries and others have humongous amounts of data of their citizens, users, customers, patients which can be used to train some very intelligent models for specific purposes. However, use of these datasets are mostly restricted because of privacy concerns about user's data. Models which use these private datasets for training tend to overfit and leak information the training dataset \cite{DBLP:conf/csfw/YeomGFJ18}. To overcome this problem, the concept of differential privacy \cite{DBLP:journals/corr/AbadiCGMMTZ16} was introduced. The main objective of differential privacy is to protect the leakage of any information about the training dataset. Differential Privacy can be applied at the three different stages of a deep learning model: input data, model training and output data. In context of our paper, we'll be considering models that are trained with differential privacy such that the model's tuned hyper-parameters don't leak any information about the training dataset. This is usually done by adding noise to the model's gradients while training. This noise does affect the accuracy of the model hence a trade-off between accuracy and privacy needs to be made depending upon the expected capability of the model.

% \begin{figure}[!t]
% % \centerline{\includesvg[width=0.75\textwidth]{TimingLeakage.drawio.svg}}
% \centerline{\includegraphics[width=0.75\linewidth]{DP_Example.drawio.pdf}}
% % \centerline{\includesvg[width=4cm]{TimingLeakage.drawio.svg}}
% \caption{Concept of Differential Privacy}
% \label{DifferentialPrivacy}
% \end{figure}

\section{Implementation Vulnerability in PyTorch}
\label{section:vulnerability}
The Python-based Deep Learning (DL) library PyTorch, developed by Meta (Facebook) AI Research, has recently found a growing interest in several industry-standard AI-enabled products because of its dynamic graph creation, data parallelism debugging, and developer-friendliness. The popularity of PyTorch even pushed several organizations to replace famous DL stacks like TensorFlow with PyTorch as a core module in various applications~\cite{sameer_2020,scott_2020}. Even though PyTorch is a prevalent and powerful library, we have identified a data-dependent vulnerability that can leak class-label information of inputs through the timing side-channel. The timing side-channel can lead to catastrophic privacy violations to the user base of a Machine Learning as a Service (MLaaS) provider, which we have discussed later in Section~\ref{section:MLP}. In this section, we first show the timing leakage vulnerability in PyTorch-based Convolution Neural Networks (CNN). Then we identify, demonstrate and analyze the source of timing leakage, which is the Max Pooling implementation of CNNs. In the following subsection, we provide details on the experimental scenario and basic setup we used for our analysis performed in this section.

\subsection{Experimental Scenario and Setup}\label{sec:exp_setup}
\begin{figure}[!b]
\centerline{\includegraphics[width=0.7\linewidth]{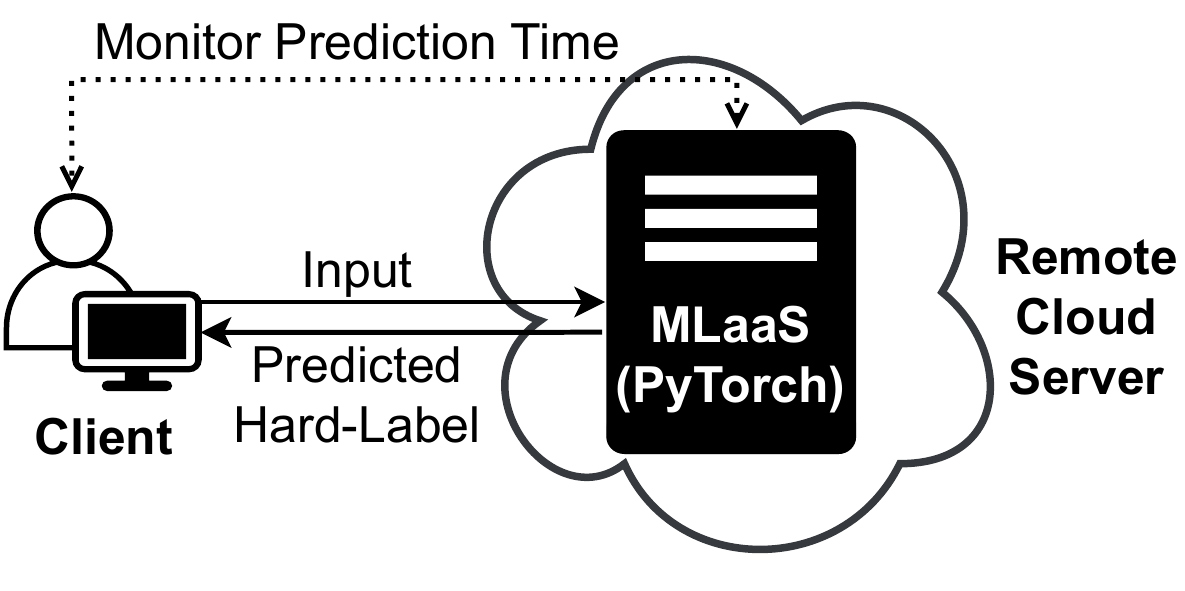}}
\caption{\textbf{Experimental Scenario:} A client having hard-label black-box access to a remote cloud-server providing MLaaS through a CNN implemented using PyTorch. The client can also monitor execution time during inference operation\vspace{-0.3cm}}
\label{Client_Server}
\end{figure}

% In this section we first discuss the experimental scenario (Figure \ref{Client_Server}), based on which we've performed all our experiments and then discuss about the experimental setup. There is remote cloud server, providing the MLaaS service. The DL model on the MLaaS has been implemented using the PyTorch library. A client on the server has black-box access to the DL model, to which it can give the input data and receive classification output. The client also has the capability to get the inference time of the input provided to the model.
\textbf{Scenario:} We consider a remote cloud server that provides MLaaS to its clients, and the cloud server uses PyTorch to implement its CNN model. We assume that a client has hard-label black-box access to CNN, i.e., the client can only query CNN with input and obtain classification output as a hard-label. In hard-label black-box access, a client can only obtain actual labels and does not get any knowledge of probabilities associated with predicted labels. We also assume that the client has the capability to monitor execution times during the inference operation using CNN. The scenario is briefly illustrated in Fig.~\ref{Client_Server}.

% \subsection{Experimental Setup}
% Now, we introduce our experimental setup for our future experiments. We have performed all our experiments on an Intel Xeon (4 cores, Skylake architecture) machine with 16GB RAM. To confirm the consistency of our results we also performed experiments on an Intel i7-4790 processor (4 cores, Haswell micro-architecture) processor system. We first establish the timing side-channel leakage on a custom Convolution Neural Network (CNN) architecture implemented using Facebook's PyTorch (1.9.1+cpu) library with Python3. The architecture of the custom CNN is provided in Table~\ref{table:custCNN}. We also support the claim using several state-of-the-art pre-trained models implemented using PyTorch, namely \textit{AlexNet}~\cite{DBLP:conf/nips/KrizhevskySH12}, \textit{DenseNet121}~\cite{DBLP:conf/cvpr/HuangLMW17}, \textit{SqueezeNet}~\cite{DBLP:journals/corr/IandolaMAHDK16}, \textit{ResNet50}~\cite{DBLP:conf/cvpr/HeZRS16}, and \textit{VGG19}~\cite{DBLP:journals/corr/SimonyanZ14a}
% for the experiments. We trained all our models over two widely-used image classification benchmark datasets for our experiments, \textit{CIFAR10}~\cite{cifar10} and \textit{CIFAR100}~\cite{cifar100}. Now, from next section we start the discussion on analysing DL timing side-channels leaking class-labels.
\textbf{Setup:} In order to first establish the timing side-channel leakage, we consider a custom CNN architecture implemented using PyTorch (1.9.1+cpu). The architecture of custom CNN is provided in Table~\ref{table:custCNN}. We also analyze several state-of-the-art pre-trained CNNs implemented using PyTorch, namely \textit{AlexNet}, \textit{DenseNet121}, \textit{SqueezeNet}, \textit{ResNet50}, and \textit{VGG19}, to support the claim of timing side-channel leakage. We consider two widely-used standard image classification benchmarking datasets, \textit{CIFAR10} and \textit{CIFAR100}, for the evaluation. In order to validate the generalizability of timing side-channel leakage in multiple computing environments, we perform all our experiments on an \textit{Intel Xeon} (4 cores, Skylake micro-architecture) machine with 16GB RAM and an \textit{Intel i7-4790} (4 cores, Haswell micro-architecture) machine with 16GB RAM. In the following subsection, we provide an overview of the timing measurement strategy and details on analyzing the timing.

\begin{table}[!t]
  \begin{center}
  \caption{Custom CNN Architecture for CIFAR10 and CIFAR100. The filter size of Convolution layers and the number of neurons in Fully Connected layers are given after layer names. MaxPool's Kernel size = 3x3 and Stride = 2}
  \begin{tabular}{cc}
    \toprule
    \textbf{Layer Name} & \textbf{Layer Type} \\
    \midrule
    Input Layer & Input\\
    Layer1 & Convolution-16\\
    Layer2 & Convolution-32\\
    Layer3 & MaxPool\\
    Layer4 & Convolution-32\\
    Layer5 & Convolution-32\\
    Layer6 & MaxPool\\
    Layer7 & Convolution-64\\
    Layer8 & Convolution-128\\
    Layer9 & Fully connected - 128\\
    Layer10 & Fully connected - 64\\
    Output Layer & Softmax\\
  \bottomrule
\end{tabular}
\label{table:custCNN}
\end{center}
\end{table}

% \subsection{Timing Measurement Setup}
% We use  \texttt{perf\_counter()} from python's \texttt{time} library to calculate the execution time of a model's prediction function. Following is a code snippet that returns the total execution time of \emph{forward} propagation, which is also PyTorch's inference operation, for a sample PyTorch model.

% % \begin{verbatim}
% \begin{lstlisting}[language=Python, caption=Time measurement of prediction function code sample]
% def forward(self,x):
%     t1 = time.perf_counter()
%     x = self.conv1(x) #Convolution
%     x = torch.nn.functional.relu(x) #ReLU activation
%     x = self.conv2(x)
%     x = torch.nn.functional.relu(x)
%     x = self.pool(x) #MaxPool
%     ...
%     x = x.view(x.size(0), -1) #Flattening
%     x = self.F1(x) #Dense Layer Operation
%     output = self.out(x)
%     t2 = time.perf_counter()
%     return output, t2-t1
% \end{lstlisting}

\subsection{Analysis of Timing Measurements}
\label{section:ATM}
% In this section, we first introduce the tools to get inference times of the inputs to the DL and then discuss that on what basis we can say a class input is distinguishable from other class input using the calculated inference time of the inputs.
% We use  \texttt{perf\_counter()} from python's \texttt{time} library to get the inference time of a DL model's prediction for a particular input. User does not require root access to run this library's functions. Following is a code snippet that returns the total execution time of \emph{forward} propagation, which is also PyTorch's inference operation, for a sample PyTorch model.
We use the \texttt{perf\_counter()} method from Python's \texttt{time} library to obtain execution time of a CNN during its inference operation for a particular input. The \texttt{perf\_counter()} method can be invoked using user-level privilege. The code snippet in Listing~\ref{lst:overall_prediction} is a sample example of obtaining the total execution time of a \textit{forward propagation} (i.e., inference operation of PyTorch library) for a sample CNN model. The term $t2 - t1$ provides the overall inference time for input~$x$.

\begin{lstlisting}[language=Python, caption=Sample code to get inference time during prediction,label={lst:overall_prediction}]
def forward(self,x):
    t1 = time.perf_counter() #Timestamp 0
    x = self.conv1(x) #Convolution
    x = torch.nn.functional.relu(x) #ReLU activation
    x = self.conv2(x)
    x = torch.nn.functional.relu(x)
    x = self.pool(x) #MaxPool
    ...
    x = x.view(x.size(0), -1) #Flattening
    x = self.F1(x) #Dense Layer Operation
    output = self.out(x)
    t2 = time.perf_counter() #Timestamp 1
    return output, t2-t1
\end{lstlisting}

% Now, we look into the criteria for deciding distinguishable  class-labels based on inference times of their inputs.
Let the inference time for any input $k$ of class $\mathcal{C}_i$ be denoted as $t_{\mathcal{C}_{ik}}$. We observe $t_{\mathcal{C}_{ik}}$ for $\mathcal{N}$ repetitions to obtain a distribution $\mathcal{T}_{\mathcal{C}_{ik}} = \{t_{\mathcal{C}_{ik}}^1, t_{\mathcal{C}_{ik}}^2, \dots, t_{\mathcal{C}_{ik}}^\mathcal{N}\}$, where $t_{\mathcal{C}_{ik}}^r$ is the value of $t_{\mathcal{C}_{ik}}$ at $r$-th repetition. We repeat the process for $\mathcal{P}$ different input examples of class $\mathcal{C}_i$ to obtain the \textit{timing distribution} $\mathcal{T}_{\mathcal{C}_i} = \{\mathcal{T}_{\mathcal{C}_{i1}} \mathbin\Vert \mathcal{T}_{\mathcal{C}_{i2}}\mathbin\Vert \dots \mathbin\Vert \mathcal{T}_{\mathcal{C}_{i\mathcal{P}}}\}$, where $\mathbin\Vert$ is the append operation and $|\mathcal{T}_{\mathcal{C}_i}| = \mathcal{P}\mathcal{N}$. We obtain $\mathcal{T}_{\mathcal{C}_i}$ for each class $i$ in the dataset and repeat the process for $\mathcal{M}$ independent runs at different time instances for a generalized analysis under different execution environment state. Let the timing distribution of class $\mathcal{C}_i$ at time instance $m$ is denoted as $\mathcal{T}_{\mathcal{C}_i}^{(m)}$. We report that an input of class $\mathcal{C}_i$ is distinguishable from an input of class $\mathcal{C}_j$ at time instance $m$, if $\tilde{\mathcal{T}}_{C_i}^{(m)} > \tilde{\mathcal{T}}_{C_j}^{(m)}$ or $\tilde{\mathcal{T}}_{C_i}^{(m)} < \tilde{\mathcal{T}}_{C_j}^{(m)}$, where $\tilde{\mathcal{T}}_{C_i}^{(m)}$ is the median of $\mathcal{P}\mathcal{N}$ values in $\mathcal{T}_{C_i}^{(m)}$. We define an indicator variable $\mathcal{B}_{m}(\mathcal{C}_i, \mathcal{C}_j)$ for the class pair $\mathcal{C}_i$ and $\mathcal{C}_j$ at $m$-th time instance as:
\begin{equation}
    \mathcal{B}_m(\mathcal{C}_i, \mathcal{C}_j) = 
    \begin{cases}
        1, &\text{if } \tilde{\mathcal{T}}_{C_i}^{(m)} > \tilde{\mathcal{T}}_{C_j}^{(m)} \\
        0, &\text{if } \tilde{\mathcal{T}}_{C_i}^{(m)} < \tilde{\mathcal{T}}_{C_j}^{(m)} 
    \end{cases}
\end{equation}
\noindent Next, we define a decision making variable $\mathcal{D}(\mathcal{C}_i$, $\mathcal{C}_j$) for the class pair $\mathcal{C}_i$ and $\mathcal{C}_j$ as:
\begin{equation}
\mathcal{D}(\mathcal{C}_i, \mathcal{C}_j) = \sum_{m=1}^{\mathcal{M}} \mathcal{B}_{m}(\mathcal{C}_i, \mathcal{C}_j)
\end{equation}
where $\mathcal{D}(\mathcal{C}_i, \mathcal{C}_j) \in [0,\mathcal{M}]$. We denote that the class pair $\mathcal{C}_i$ and $\mathcal{C}_j$ is distinguishable using the inference time from CNN, in general, if
\begin{center}
$0 \le \mathcal{D}(\mathcal{C}_i, \mathcal{C}_j) \le \frac{\mathcal{M}}{2}-2$ or $\frac{\mathcal{M}}{2}+2 \le \mathcal{D}(\mathcal{C}_i, \mathcal{C}_j) \le \mathcal{M}$
\end{center}
else we consider the pair to be indistinguishable. The class pairs for which $\frac{\mathcal{M}}{2}-1 \le \mathcal{D}(\mathcal{C}_i, \mathcal{C}_j) \le \frac{\mathcal{M}}{2}+1$, we can not conclude that the inference time of one class is greater than the other in general. For such pairs, $\mathcal{D}(\mathcal{C}_i, \mathcal{C}_j)$ lies near the mean of $\mathcal{M}$, suggesting the inference times for both classes are, in general, approximately similar.

% For the pairs whose $D$ value lies in the critical range $[\frac{M}{2}-1,\frac{M}{2}+1]$, that is the range inside $10\%$ variation of the mean of $M$,  
%  Hence, the $D$ value in this range suggests that timing values for both classes are very similar and hence hard to distinguish. In our experimental setup we have chosen $P=10$, $N=100$ and $M=10$.

\subsubsection{Inference Time Analysis}
\label{section:ATM_PTA}
% We begin by analyzing the prediction time of the customized CNN model by training it over the CIFAR10 dataset. Our first step was to do a timing experiment over the prediction function for this model and check whether there is any difference between the inference time for different classes of input data. We have ten classes in the CIFAR10 dataset, hence a total of ${10\choose 2}=45$ class pairs. We observed that  43 out of the 45 class pairs were distinguishable based on the inference time, which is quite high, and hence indicate towards timing leakage. Now, to discover the underlying cause of this timing difference we do a layer-wise analysis experiment in the next section.
% \par We got results of the above experiments for five other statemodels as well Alexnet, Squeezenet, Densenet, VGG19, and Resnet50. The results are shown in Figure~\ref{fig:distinguish} for CIFAR10 and CIFAR100. In CIFAR100, there are 100 classes, hence a total of ${100\choose 2}=4950$ class pairs. We observed that for all these models with the CIFAR10 and CIFAR100 dataset, we could distinguish the majority of class pairs based on timing.

\begin{figure}[!t]
    \centering
    % \subfigure[]{\includesvg[width=0.4\textwidth]{CIFAR10_100_red.svg}}
    % \subfigure[]{\includesvg[width=0.4\textwidth]{CIFAR10_100_i7_red.svg}}
    % \subfloat[\label{fig:distinguish:xeon}]{\includesvg[width=0.5\linewidth]{CIFAR10_100_red.svg}}
    \subfloat[\label{fig:distinguish:xeon}]{ \includegraphics[width=0.5\linewidth]{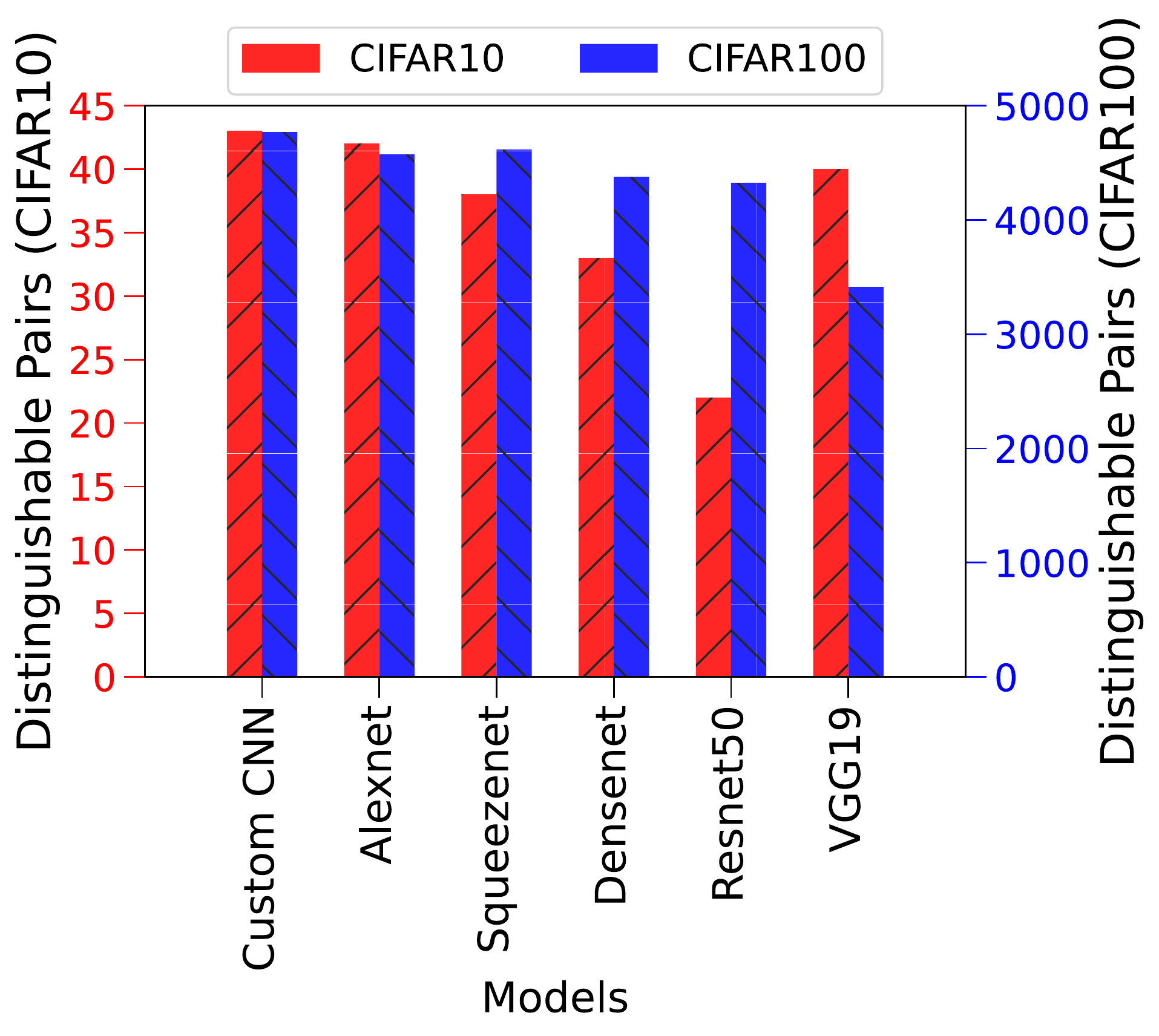}}
    % \subfloat[\label{fig:distinguish:i7}]{\includesvg[width=0.5\linewidth]{CIFAR10_100_i7_red.svg}}
    \subfloat[\label{fig:distinguish:i7}]{\includegraphics[width=0.5\linewidth]{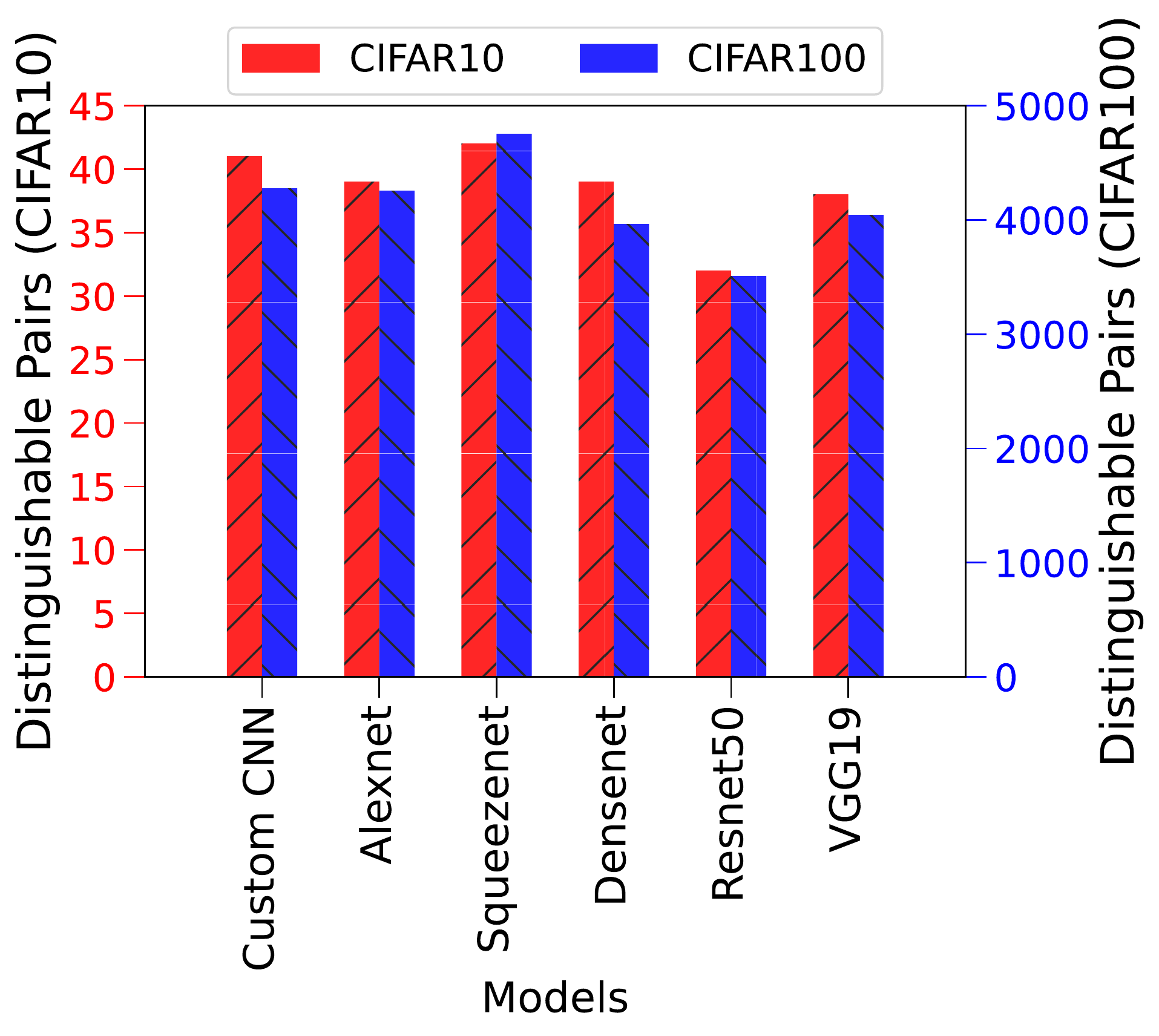}}
    \caption{Number of distinguishable class pairs using timing side-channel in different CNN models on CIFAR10 (out of 45) and CIFAR100 (out of 4950) on (a) Intel Xeon and (b) Intel i7 machines\vspace{-0.4cm}}
    \label{fig:distinguish}
\end{figure}

% \par \textbf{Results with Intel i7: } To confirm that the observed time difference is not processor specific, we performed the same experiment on an Intel i7-4790 processor (4 cores, Haswell micro-architecture). The results are shown in Figure \ref{fig:distinguish}.  They are very similar to what we observed before, majority of the class pairs are distinguishable for both CIFAR10 and CIFAR100 with all models.
We perform the inference time analysis, as discussed above, for the custom CNN mentioned in Table~\ref{table:custCNN}. We observe that out of 45 class pairs in CIFAR-10, 43 class pairs can be distinguished, and out of 4950 class pairs in CIFAR-100, 4767 class pairs can be distinguished based on the inference time in Intel Xeon machine\footnote{CIFAR-10 has images of 10 classes indicating a total of ${10\choose 2}=45$ class pairs, and CIFAR-100 has images of 100 classes indicating a total of ${100\choose 2}=4950$ class pairs.}. In Intel i7 machine, the results for CIFAR-10 and CIFAR-100 are 41 and 4273, respectively. The high number of distinguishable pairs for both datasets indicates data-dependent timing-leakage, generalized over different computing environments. We perform the same analysis on five different state-of-the-art pre-trained CNN models, as discussed in Section~\ref{sec:exp_setup}, to investigate the existence of data-dependent timing side-channel leakage in other CNNs. The results of inference time analysis for all these models considering CIFAR-10 and CIFAR-100 datasets for both Intel Xeon and Intel i7 machines are shown in Fig.~\ref{fig:distinguish:xeon} and Fig.~\ref{fig:distinguish:i7}, respectively. The vertical axis in the figure represents the total number of distinguishable class pairs for CIFAR-10 (red) and CIFAR-100 (blue). We can observe that most class pairs in CIFAR-10 and CIFAR-100 can be distinguished in all these models using only the inference time.

\subsubsection{Layer-wise Inference Time Analysis}
\label{section:ATM_LPTA}
% To further investigate the reason behind the timing difference, we did a layer-wise timing experiment on the custom CNN. We first calculated the execution time of each layer for different class inputs inputs. We treat activation functions as a separate layer. Following is a code snippet for a sample model to get layer-wise timing values during the inference operation:
In order to investigate the source behind the existence of data-dependent timing side-channel leakage, as demonstrated in the previous subsection, we perform the same analysis using the execution time of each layer during the inference phase. The code snippet in Listing~\ref{lst:layer_prediction} is a sample example of obtaining layer-wise execution time during the inference operation of a sample CNN model.

\begin{lstlisting}[language=Python, caption=Sample code to get execution time of each layer during prediction,label={lst:layer_prediction}]
def forward(self,x):
    t=[]
    t.append(time.perf_counter()) #Timestamp 0
    x = self.conv1(x) #Convolution
    t.append(time.perf_counter()) #Timestamp 1
    x = torch.nn.functional.relu(x) #ReLU activation
    t.append(time.perf_counter()) #Timestamp 2
    x = self.conv1(x)
    t.append(time.perf_counter()) #Timestamp 3
    x = torch.nn.functional.relu(x)
    t.append(time.perf_counter()) #Timestamp 4
    x = self.pool(x) #MaxPool
    t.append(time.perf_counter()) #Timestamp 5
    ...
    x = x.view(x.size(0), -1) #Flattening
    t.append(time.perf_counter()) #Timestamp 18
    x = self.F1(x)  #Dense Layer Operation
    t.append(time.perf_counter()) #Timestamp 19
    output = self.out(x)
    t.append(time.perf_counter()) #Timestamp 20
    return output, t
\end{lstlisting}

\begin{figure}[!ht]
    \centering
    % \subfigure[]{\includegraphics[width=0.35\textwidth]{ReLU_Layerwise_python_solid.png}} 
    % \subfigure[]{\includegraphics[width=0.35\textwidth]{AvgPool_ReLU_Layerwise_python_solid.png}} 
    \subfloat[\label{fig:leakage_max_pool}]{\includegraphics[width=0.4\linewidth]{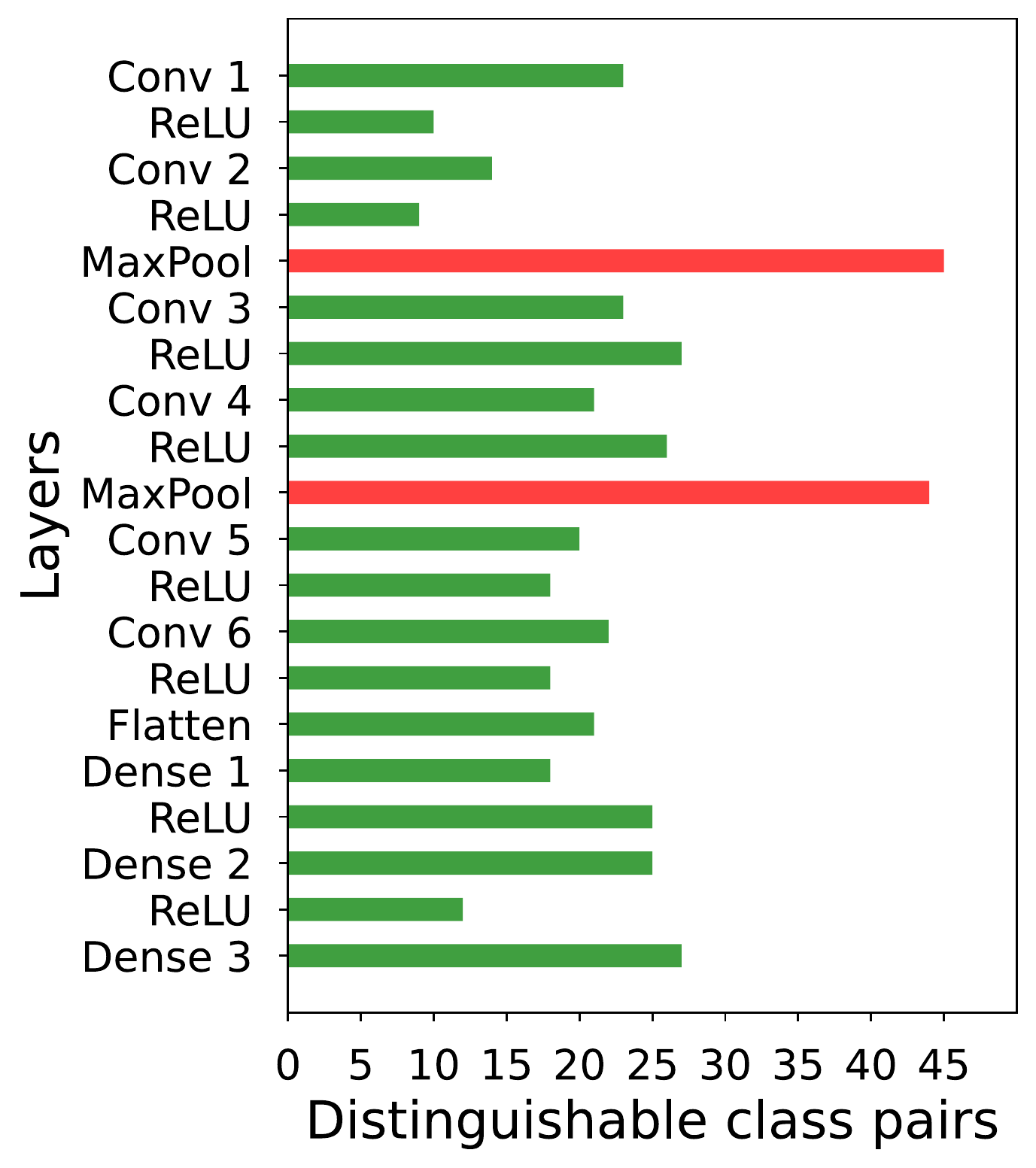}}\qquad
    \subfloat[\label{fig:leakage_avg_pool}]{\includegraphics[width=0.4\linewidth]{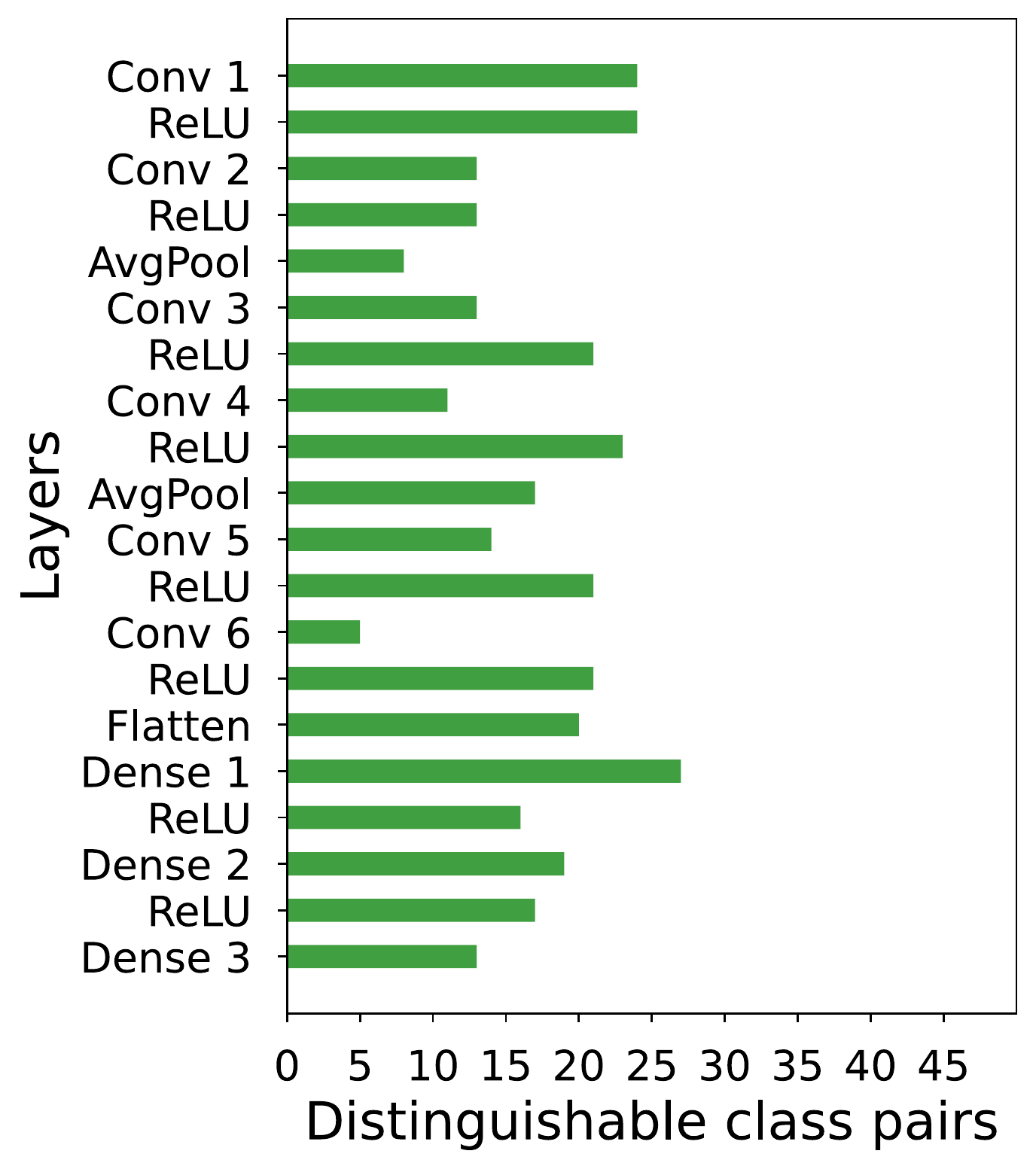}}\qquad
    \caption{Number of distinguishable class pairs (out of 45) using timing side-channel of each layer in Custom CNN on CIFAR10 with (a) Max Pooling (b) Average Pooling}
    \label{fig:fig_layers}
\end{figure}

% In the code, we get the time stamps after execution of each layer inside the \emph{forward} function and then return the final array $t$ containing timestamps for all layers.
% From array $t$, we easily compute the time for all layers by subtracting adjacent values.
% In total, we had 20 timing values for one run of the prediction. These 20 timings were for convolution, activation, pooling, and dense layer operation. We did a similar median analysis as we discussed in subsection \ref{section:ATM} to get the distinguishable class pairs for all layers. The results are shown in Figure~\ref{fig:fig_layers}. The results of this experiment showed that all layers could distinguish some of the input class pairs, but the pooling function/layer was able to distinguish the majority of the pairs and is the primary contributor to the observed differences in inference time. 
We accumulate different timestamps into variable $t$ after the execution of each layer during forward propagation. From the final values in $t$, we can compute the execution time of each layer by subtracting adjacent values. Without loss of generality, we perform inference time analysis on custom CNN considering CIFAR-10 on Intel Xeon machine to obtain total number of distinguishable class pairs using execution times of each layer. The result of the analysis is shown in Fig.~\ref{fig:leakage_max_pool}. The horizontal axis in the figure represents the total number of distinguishable class pairs for each layer represented in the vertical axis\footnote{We consider activation functions as separate layers.}. We can observe that all layers can distinguish different numbers of class pairs. However, the Max Pooling layer can distinguish the most number of pairs and is considered to be the primary contributor to observed timing differences using the overall inference time.

\subsubsection{Timing Analysis with Average Pool}
Average Pooling\footnote{Average Pooling reduces the dimension of feature maps by averaging out the values in a sliding window fashion.} is another very commonly used pooling function in CNN architectures, hence we try to verify whether this vulnerability is limited to max pooling or found in other pooling functions as well.  We repeat the layer-wise timing experiment by replacing all MaxPools in the Custom CNN model with Average Pool. The results are given in Fig.~\ref{fig:leakage_avg_pool}. We observed that whereas the max-pooling layers could completely distinguish all class pairs, the average pooling layer was only able to distinguish less than fifty percent of the class pairs. Hence, we can say this vulnerability is not present in the Average pooling function and hence we further explore the cause behind the leakage in Maxpool.
% This is because the `if' condition which was present in the PyTorch's MaxPool implementation to get the maximum value of a window will not be required in the implementation of Average Pool. All the elements of a window will be used to calculate the average of a window, and the computation will be similar for all the inputs to the CNN.

\subsection{Analysis on PyTorch Maxpool Implementation}
\label{section:pytorch_vul}
\begin{figure}[!t]
\centerline{\includegraphics[width=0.3\textwidth]{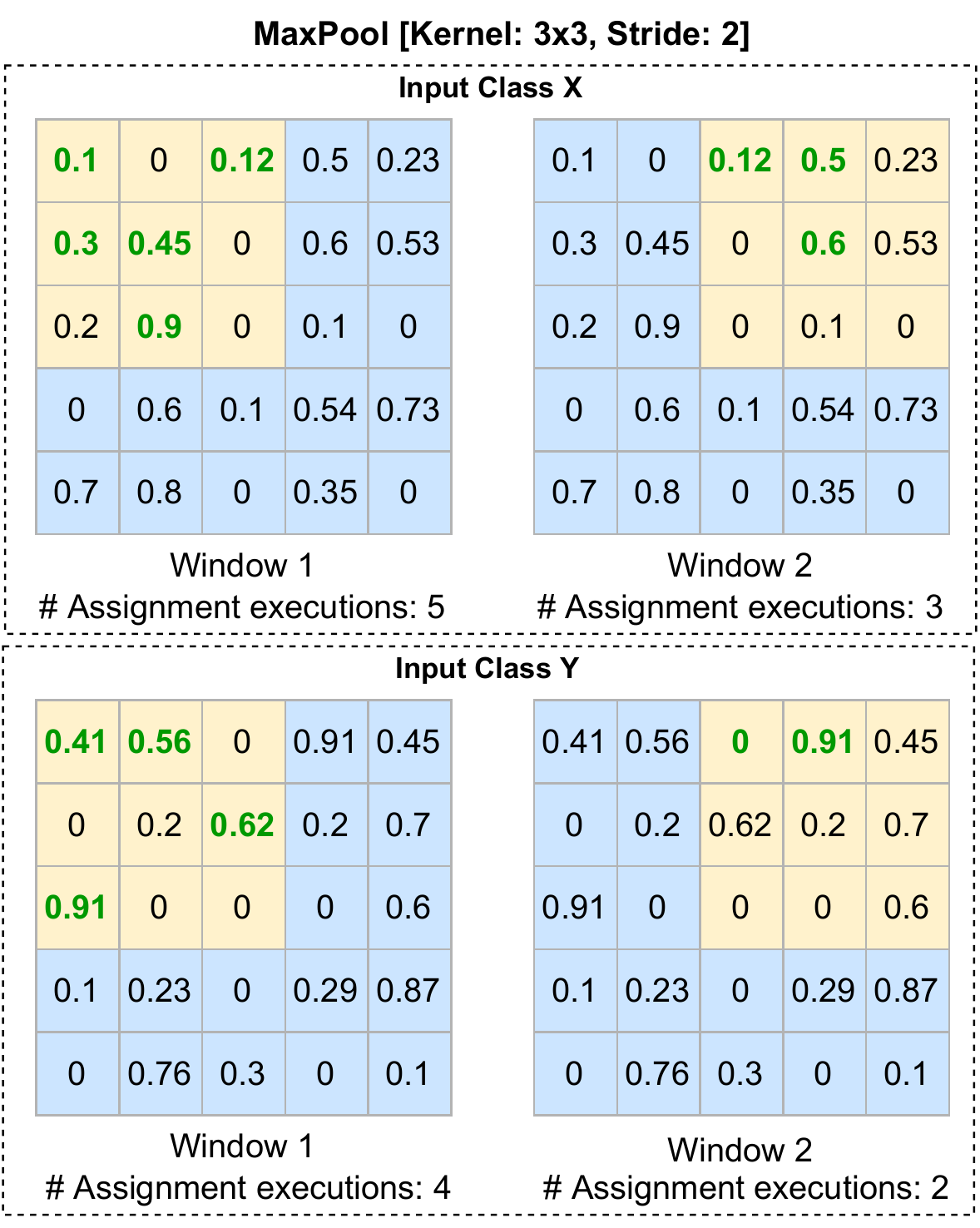}}
\caption[An example of varying number of if statement calls for two different windows (Kernels) during Maxpool operation]{An example of varying number of if statement calls for two different windows (Kernels) during Maxpool operation}
\label{if_stat_eg}
\end{figure}

The MaxPool function slides through defined kernel size matrices to get reduced feature maps as seen in Fig.~\ref{maxpool}. Ideally, this function should take constant-time for all inputs, but our results say otherwise. Hence, we look into the PyTorch implementation of the Maxpool function to closely analyze the cause of the vulnerability.  To get the maximum value for each pooling window, an `if' condition is used, which checks all elements and keeps updating the max value when it finds a greater value. It also updates the index of the current max value. The code snippet from PyTorch Github repository\footnote{PyTorch Github repository code snippet (line 65-68):\url{https://github.com/pytorch/pytorch/blob/bceb1db885cafa87fe8d037d8f22ae9649a1bba0/aten/src/ATen/native/cpu/MaxPoolKernel.cpp\#L65}} is shown in Listing~\ref{code_snippet}.
\begin{lstlisting}[language=Python, caption=PyTorch Maxpool’s code snippet to update max, label=code_snippet]
if ((val > maxval) || std::isnan(val)) {
    maxval = val;
    maxindex = index;
}
\end{lstlisting}
Here $val$ is the value of the element at current $index$ and $maxval$ stores the value of the maximum element found till now in the current window. For each window, the number of times assignment statement inside if statement is executed depends on the position of the max value in the window. Hence, the overall number of assignment statements executed inside the `if' statement differs for different inputs to Maxpool. This is illustrated in Fig.~\ref{if_stat_eg} using an example. The yellow $3 \times 3$ windows represents the current window of the input matrix on which the Maxpool is being operated. The bold green text emphasizes on the indices of the matrix, for which the assignment operation inside the `if' statement has been executed. We see in the figure, that for two different inputs the assignment executions inside the `if' statement vary in both kernel-sized windows of both the inputs. The total assignment executions for input $X$ and $Y$ differ from each other, hence causing difference in the execution time of the Maxpool operation and the overall inference time of an input data as well.

% To confirm whether this is the cause behind getting a difference in timing between the class pairs, we ran an experiment using our Custom CNN with the CIFAR10 dataset.

% \vspace{5mm}

\subsubsection{Vulnerability in the implementation}
Since, the number of times the assignment operations inside `if' (branch not taken) statement is executed depends on the input to the MaxPool function, it makes the implementation dependent on the input data given to the MaxPool function .The input data is resultant of multiple transformations (like convolution and activation operations) applied to the original input provided to the DL model, hence there is a correlation between the two, which indirectly makes the total number of assignment operations dependent on the input to the model
\subsubsection{Influence on Timing}
\begin{figure}[!t]
    \centering
    \subfloat[]{\includegraphics[width=0.4\linewidth]{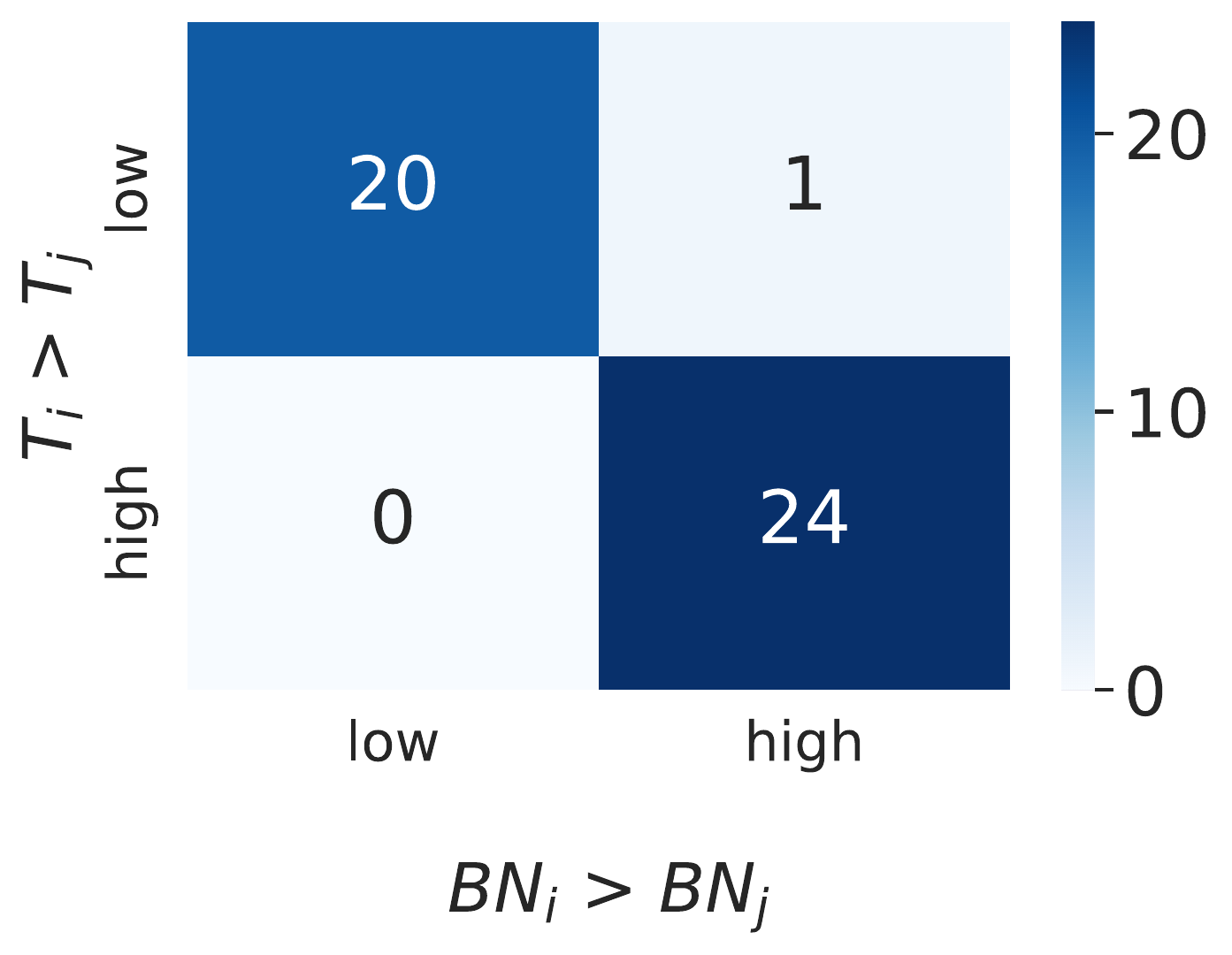}}\qquad
    \subfloat[]{\includegraphics[width=0.4\linewidth]{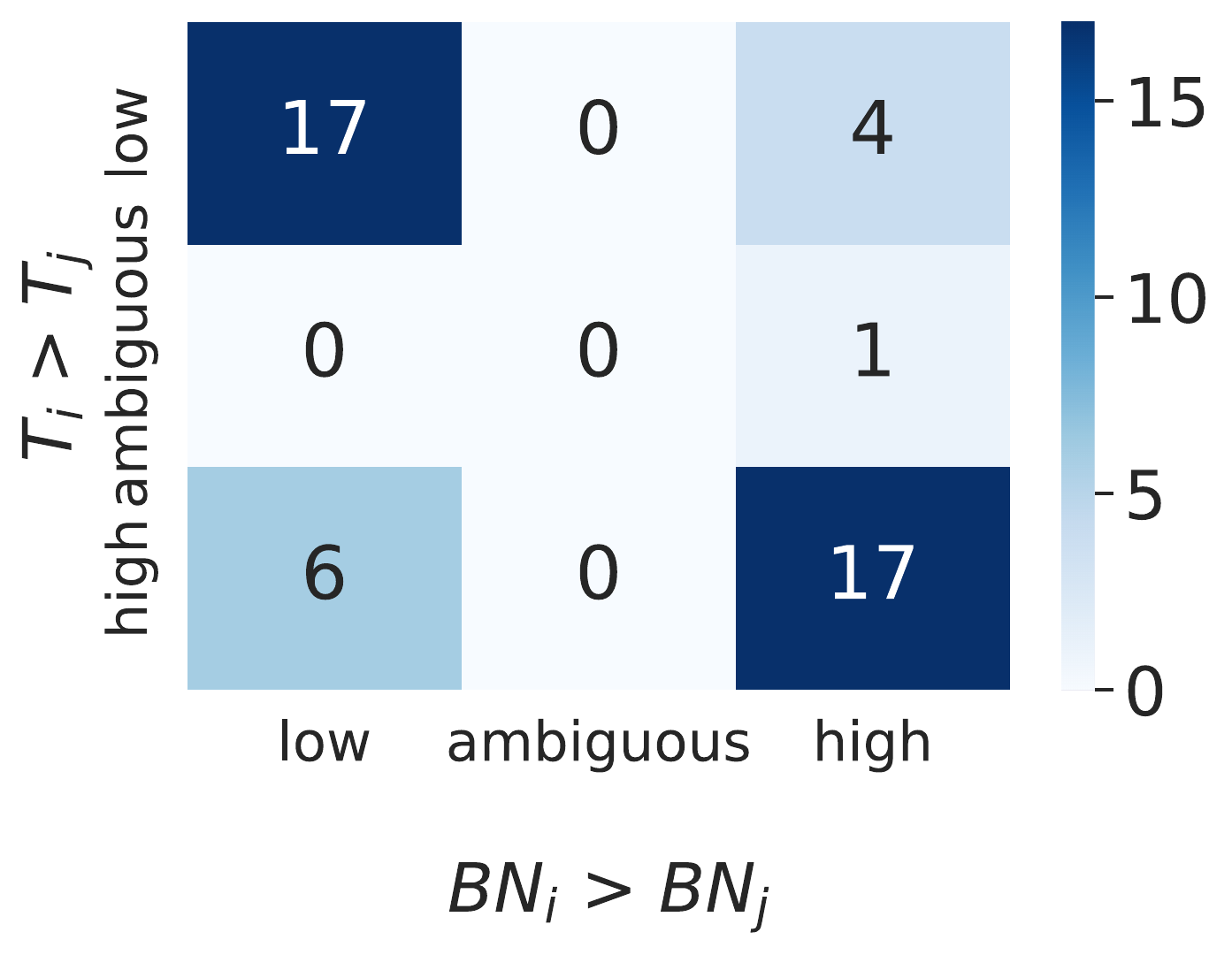}}\qquad
    
    \caption{(a) First Maxpool Layer Timing correlation with Branch not taken instructions (b) Second Maxpool Layer Timing correlation with Branch not taken instructions}
    \label{fig:correlation_BN}
\end{figure}

% \par\textbf{Influence on timing:}
In this work we show how an adversary can exploit this implementation vulnerability to know the class labels of the input data using timing side-channels. It is possible because, different class labels execute different number of `if' statements inside the MaxPool implementation. This causes a timing variation in the overall inference time for all inputs hence aiding an adversary to distinguish among them the different class labels. We verify this with experimental analysis on the custom CNN model to do the analysis which has two Maxpool layers. We get input tensors to both MaxPool functions for one image of each class. We calculated the number of `if' statement executions (branch not taken instructions) for all class images for both max pools. We compared these values for all class pairs and observed that the results were directly proportional to the timing results i.e., when the number of `if' statement executions for an input class is higher than another class, its execution time will also be higher. The results are shown in Fig.~\ref{fig:correlation_BN}(a) and \ref{fig:correlation_BN}(b). Here $T_i$ and $T_j$ are median of inference time distributions of some class pair $C_i$ and $C_j$ and $BN_i$ is the number of Branch not taken (if statements execution) instructions executed by max pool function for class $C_i$.

% \subsection{Analysis on State-of-the-art CNNs}
% \par Till now we've seen all the timing channel analysis with only the custom CNN model. To confirm, that this vulnerability can be observed across all PyTorch implemented models, we got results of the experiments discussed in Section \ref{section:ATM} for five other state of the art models as well, which are: Alexnet, Squeezenet, Densenet, VGG19, and Resnet50. The results are shown in Figure~\ref{fig:distinguish}(a)for CIFAR10 and CIFAR100. In CIFAR100, there are 100 classes, hence a total of ${100\choose 2}=4950$ class pairs. We observed that for all these models with the CIFAR10 and CIFAR100 dataset, we could distinguish the majority of class pairs based on timing.

% \par \textbf{Results with Intel i7: } To further confirm that the observed time difference is neither architecture specific nor processor specific, we did the overall inference time analysis on an Intel i7-4790 processor (4 cores, Haswell micro-architecture) as well. The results are shown in Figure \ref{fig:distinguish}(b).  They are very similar to what we observed before, majority of the class pairs are distinguishable for both CIFAR10 and CIFAR100 with all models.

In the next section, we discuss on a practical threat model where an adversary can exploit the class-leakage to compromise user privacy.

\section{Timing based MLP class-label classifier}
\label{section:MLP_Attack}
In previous section we identified the primary component responsible for timing leakage in PyTorch library and the effect of such leakages is visible over several popular datasets. But this non-constant timing behavior is capable of an end-to-end attack demonstration which we are going to discuss next. For this we need a standard vulnerability analysis by first defining a threat model and then based on that demonstrate an attack to infer class-labels using timing channels.
\subsection{Threat Model}
\label{section:threat_model}
\begin{figure}[!t]
% \centerline{\includegraphics[width=0.45\textwidth]{ThreatModel_multiple_clients.png}}
\centerline{\includegraphics[width=0.45\textwidth]{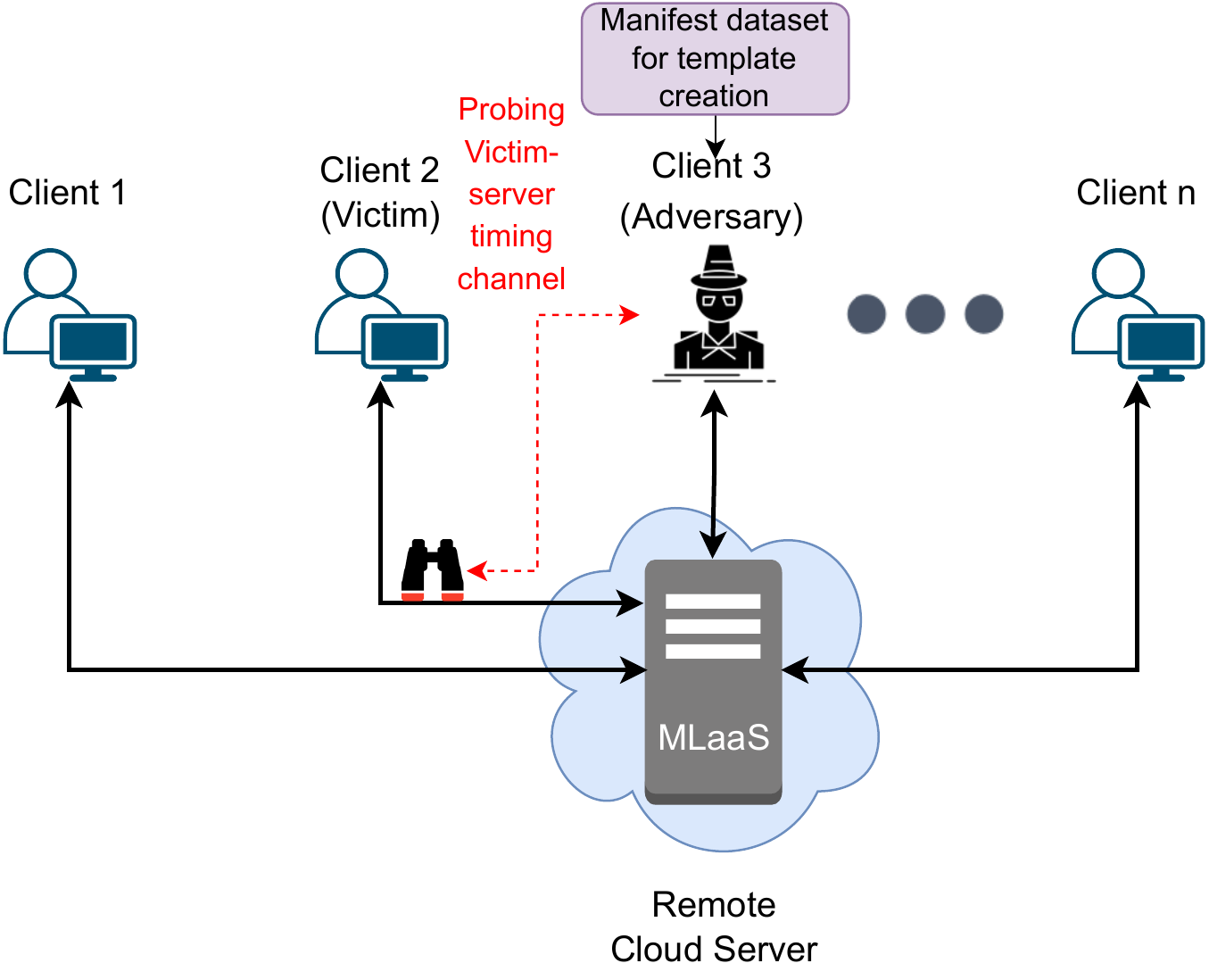}}
\caption[Threat Model]{Threat Model for Privacy Leakage in a Client-Server MLaaS Framework}
\label{threat_model}
\end{figure}

We consider a scenario where multiple clients are connected to a trusted cloud server providing Machine Learning as a Service (MLaaS) (Fig.~\ref{threat_model}. 
% The model network is trained by one of the clients using his own private dataset. He uses Differential Privacy to keep the training data privacy intact. 
Using this service, the clients can provide their private/confidential data to the MLaaS to get back classification results for a particular task. Any of these clients can be an adversary if he/she has any malicious intention of knowing the private data of another client provided to the MLaaS. 
\subsubsection*{Adversary's Capability and Objective}
The adversary is considered to have a hard-label black-box access to the DL model on the MLaaS server with user-level privilege. So he/she is capable of probing the communication channel between victim-client and the server and in order access the inference time of the inputs provided by the victim to the MLaaS. The adversary also has an access to a manifest dataset. Manifest dataset is a set of inputs which may or may not be the part of the training dataset but belongs to the same distribution. \emph{It is to be noted that, the adversary and the victim client co-reside on the same server and access the same deep learning model.} With these capabilities, adversary's objective is to infer class labels of input data fed by the chosen victim client to the MLaaS. In addition, the adversary is also capable of launching a profiling attack using Multi-Layer Perceptron (MLP).

\begin{figure}[!ht]
\centerline{\includegraphics[width=0.4\textwidth]{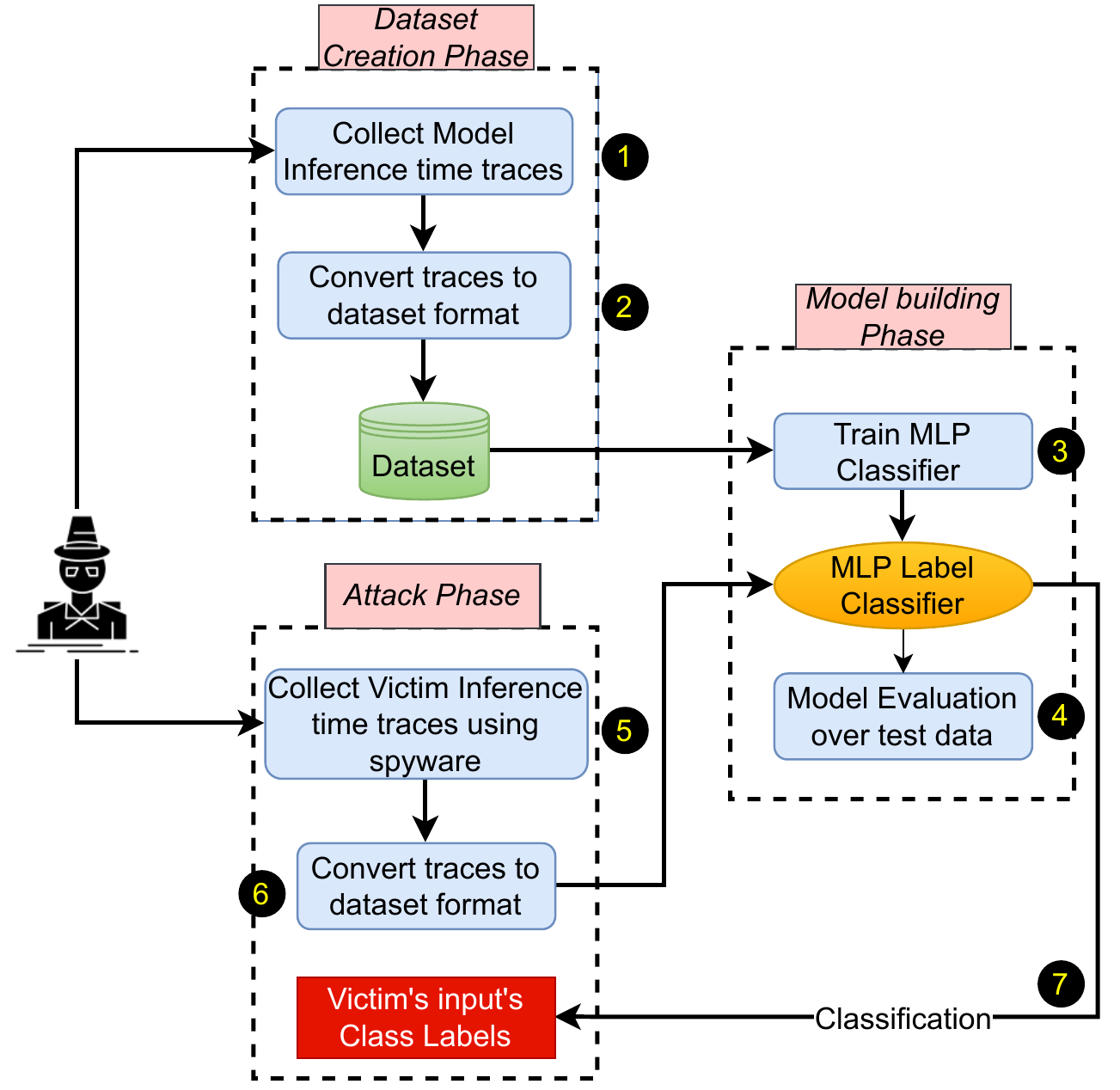}}
\caption{Attack Methodology: The process goes through three stages (i) Dataset creation Phase, (ii) Model Building Phase and (iii) Attack Phase (The numbered circles indicate the order of the process)}
\label{attack_mlp}
\end{figure}

% \section{Timing based MLP class-label classifier}
\label{section:MLP}
To confirm the viability of our results showing the vulnerability in the PyTorch library, we launched a profiling timing attack using a MLP classifier. The objective is to see whether the MLP will be able to learn the timing difference among the different classes. We divide the process in three main parts: Dataset creation, tuning and training the model, and testing the model on completely new data. The complete process flow is shown in Fig.~\ref{attack_mlp}.

\subsection{MLP Class-label Classifier Construction}

% \begin{figure*}[!t]
%     \centering
%     \subfloat[]{\includesvg[width=6.5cm]{conf_matrix_MLP.svg}}\qquad
%     \subfloat[]{\includesvg[width=6.5cm]{conf_matrix_MLP_DP.svg}}\qquad
    
%     \caption{Confusion Matrix for timing test data on the MLP model with timing values from (a) normally trained custom CNN and (b) custom CNN trained with differential privacy}
%     \label{fig:cm_MLP}
% \end{figure*}

Without loss of generality, we created a dataset of dimension $10000\times100$. Let the inference time for a particular input $k$ of class $C_i$ be denoted as $t_{C_{ik}}$. We observe inference time $t_{C_{ik}}$ for $N$ time instances to obtain a timing distribution $\mathcal{T}_{C_{ik}} = \{t_{C_{ik}}^1, t_{C_{ik}}^2, \dots, t_{C_{ik}}^N\}$ of that particular input. Now repeating this procedure over all inputs of the class, we get $\mathcal{T}_{C_{ik}}$ for $P$ different inputs of class $C_i$ to get $\mathcal{T}_{C_{i}} = \{\mathcal{T}_{C_{i1}}, \mathcal{T}_{C_{i2}}, \dots, \mathcal{T}_{C_{iP}}\}$. Next, representative timing points are generated by selecting the statistic value of the respective distributions. Medians of all P distributions in $\mathcal{T}_{C_{i}}$ are denoted as $\mathcal{M}_{C_{i}} = \{\tilde{\mathcal{T}_{C_{i1}}}, \tilde{\mathcal{T}_{C_{i2}}}, \dots, \tilde{\mathcal{T}_{C_{iP}}}\}$. This $\mathcal{M}_{C_{i}}$ array makes one row of our dataset with class label $C_i$. To avoid underfitting we make our dataset bigger by repeating the same experiment for all classes $M$ times with $P=100$, $N=500$ and $M=1000$ respectively.
\subsubsection*{Model Training}
The dataset as constructed with the representative median samples are split into training (80\%) and testing data (20\%). We use Scikit-learn's \texttt{MLPClassifier}~\cite{DBLP:journals/corr/abs-1201-0490} to build our model. Further, to get the best fit for our model, we used Scikit-learn's \texttt{GridSearchCV} functionality which takes in a set of different parameters such as, activation functions, learning rate, and network size, and then returns the set of parameters which fit the model best in terms of accuracy. Additionally, to avoid over-fitting the model we used K-fold validation method where we chose $K=10$.
\subsubsection*{Testing}
The testing phase is carried out on the twenty percent test data which weren't used while training as well as on completely new data of 2000 rows containing 200 data points for each class label. We used the new data as a confirmatory test that the model wasn't over-fitting on the split test data.

\subsection{Results and Analysis}
\label{section:Attack_CNN}
% \shubhi{This result section is very weak, please explain what do you mean by custom CNN, again the results are not explained well. Figure 11 lacks proper explanation. Please update, its hard to appreciate the results just by looking at the current writeup}

\begin{figure}[!t]
% \centerline{\includegraphics[width=0.4\textwidth]{CIFAR10_100_DP_red.png}}
\centerline{\includegraphics[width=0.6\linewidth]{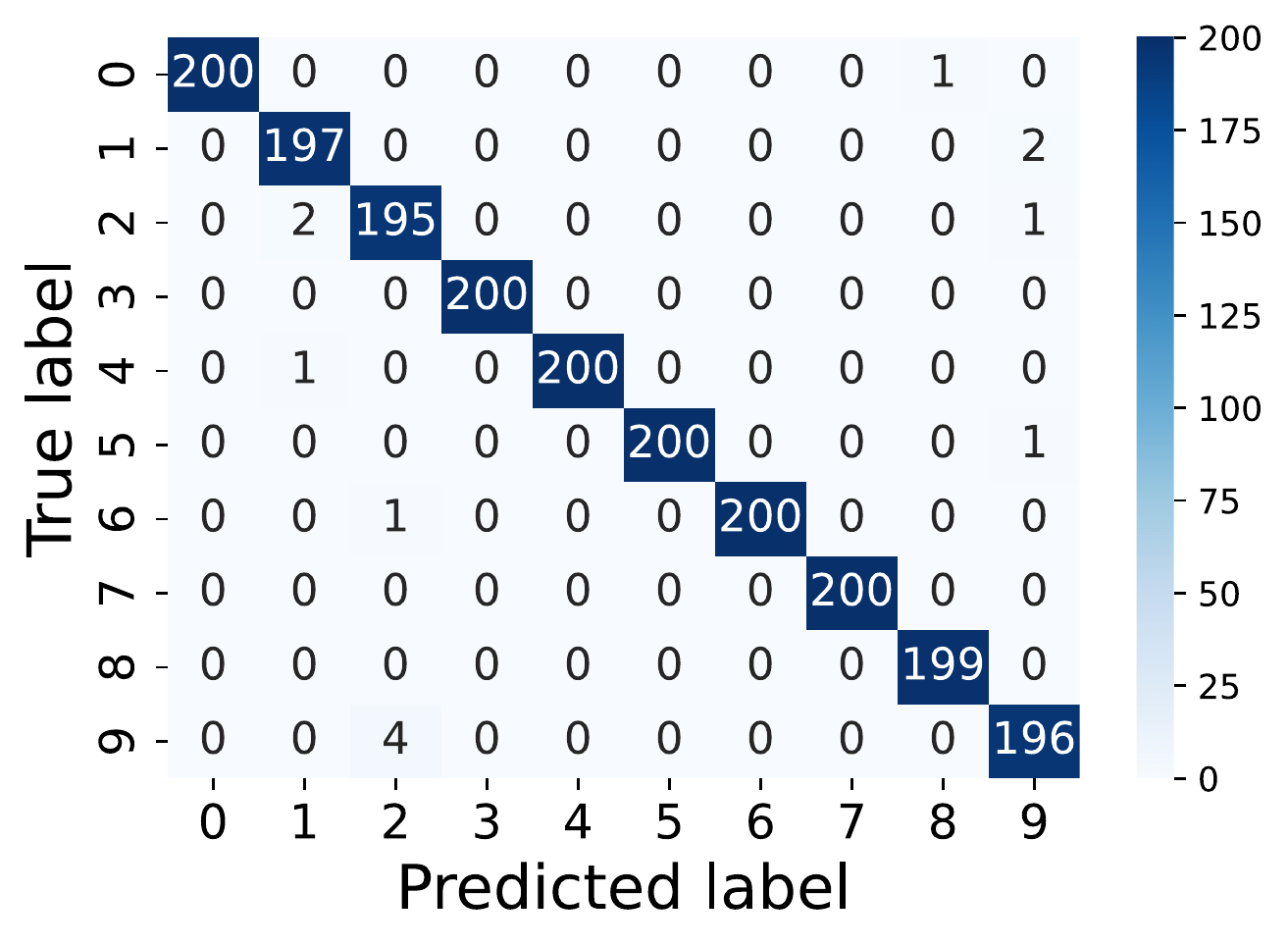}}
\caption{Confusion Matrix for timing test data on the MLP model with timing values from custom CNN}
\label{fig:cm_MLP}
\end{figure}

In this section, we give performance results of the MLP model, trained to classify class-labels using model inference time. Based on previous section, we require a  MLP classifier 10000 and 2000 ($200\times10$)
data-points for training/testing and to check over-fitting of the model on fresh data respectively. The data-points are built using processed inference time values from a CNN model for all classes. For the CNN model, we again select the custom CNN model used in Section \ref{section:ATM} as well for timing analysis of PyTorch vulnerability. (Refer Table \ref{table:custCNN} for the architecture details). We achieved a high class-label classification accuracy of approximately 99.35\% on our unused test data with 2000 inputs, fed to the MLP classifier. In Fig.~\ref{fig:cm_MLP}, we show the confusion matrix for the classification of the test data. We see that classes \emph{0, 3, 4, 5, 6,} and \emph{7} are classified with 100\% accuracy, and misclassification rate for remaining classes is less than 2.5\%. These results imply that the PyTorch vulnerability can be exploited using the proposed profiling attack and once again confirm the timing leakage as well. Next, we discuss the timing analysis results and attack scenarios in the case of differential-private models.

% ---------------------------

\section{Class-Leakage in CNN with Differential Privacy}
\label{section:DP_CL_CNN}
% In Section \ref{section:CL_CNN} we observed that when using a simple PyTorch CNN model with no additional security we can distinguish between majority of the class pairs based on their inference times. This is an ideal scenario, but in practice it's rarely the case. Model architects generally employ various security and privacy protecting mechanisms in their models. In section \ref{section:threat_model} we discussed how by knowing the class-labels of the inputs an adversary can launch a membership inference attack. Differential privacy is the most common mechanism used to prevent such an attack, in cases where the model leaks information about the training dataset. Hence, it becomes crucial to verify whether the PyTorch vulnerability still sustains, even after training the model with differential privacy.

% Protection of training data from leaking through a trained machine learning model has been a big concern recently, when companies have excessive private data about their clients. 
In Section \ref{section:ATM} we observed one aspect of PyTorch's timing vulnerability, which is its capability to leak information about the class labels. Now, in this section, we explore another possibility of a certain kind of privacy leakage caused by this vulnerability. In the introduction, we briefly discussed about membership inference attack, whose main objective is to identify whether a pair of input and output of a model, belongs to the training set used to train that model. The most common defense against this attack is differential privacy. Differential Privacy adds a certain amount of perturbation to the model hyper-parameters during the training process such that the model does not overfit on the training data. The amount of noise added to the model can be tuned using a parameter called privacy budget denoted by $\mathcal{\epsilon}$. The privacy budget is used to balance between model's performance and its privacy protection capability.

\par It would be interesting to see that, if we train our models with differential privacy, is there a way to leak information about the training data using PyTorch's vulnerability. To begin with, we first verify whether the vulnerability still persists after training our model with differential privacy, and then see how we can bypass it with the help of the attack proposed earlier.

% \subsection{Experimental Setup}
% The experimental setup is same as Section \ref{section:ATM}. The only difference is that we'll be using a library called `Opacus'~\cite{opacus} which is specifically designed to provide differential privacy to models implemented on PyTorch. Only few additional lines of code are required to convert a deep neural network(DNN) to a differentially private DNN.

\subsection{Analysis of timing vulnerability with Differential Privacy}
In this section, the experimental setup remains the same as Section \ref{section:ATM} for all experiments. In addition to that, we modify the basic Custom CNN model and five other state-of-the-art CNN models: Alexnet, Resnet50, Densenet, VGG19, and Squeezenet, by training them with differential privacy. For this purpose, we use \emph{Opacus}\footnote{Opacus Github repository: \url{https://github.com/pytorch/opacus}}~\cite{DBLP:journals/corr/abs-2109-12298} library, a research initiative by Facebook to provide differential privacy to DL models implemented on PyTorch. In the following, we begin our analysis by looking into the overall inference time for the differential-private model.

\subsubsection{Analysing Overall Inference Time}
\begin{figure}[!t]
% \centerline{\includegraphics[width=0.4\textwidth]{CIFAR10_100_DP_red.png}}
\centerline{\includegraphics[width=0.6\linewidth]{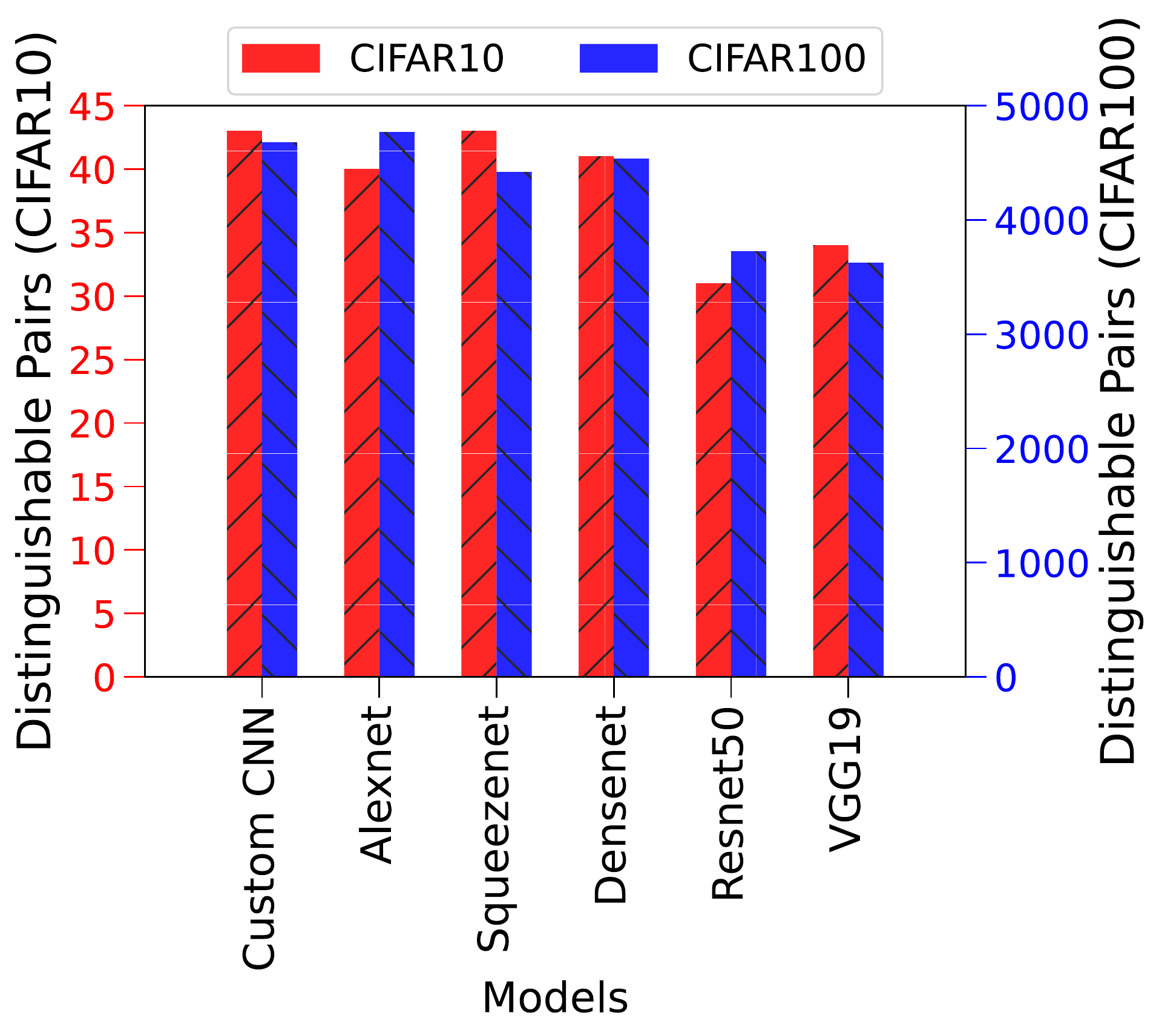}}
\caption{Number of class pairs (out of 45) distinguishable by different CNN models trained with differential privacy on CIFAR10 (out of 45) and CIFAR100 (out of 4950) dataset (Intel Xeon processor)}
\label{fig:distinguish_DP}
\end{figure}

The experiment structure remains the same from Section \ref{section:ATM_PTA}, as to analyze the effect of PyTorch vulnerability on our differential-private models. The results as illustrated in Fig.~\ref{fig:distinguish_DP} shows that the Custom CNN, Alexnet, Squeezenet, Densenet, Resnet50 and VGG19 are able to distinguish $95.55\%$, $88.8\%$, $95.55\%$, $91.11\%$, $68.8\%$ and $75.55\%$ class pairs for CIFAR-10; and $94.5\%$, $96.36\%$, $89.25\%$, $91.63\%$, $75.2\%$ and $73.2\%$ for CIFAR-100. The results confirm that the majority of the class pairs are distinguishable using the timing side-channel.

\subsubsection{Analysing Layer-wise Inference Time}
\begin{figure}[!t]
\centerline{\includegraphics[width=0.5\linewidth]{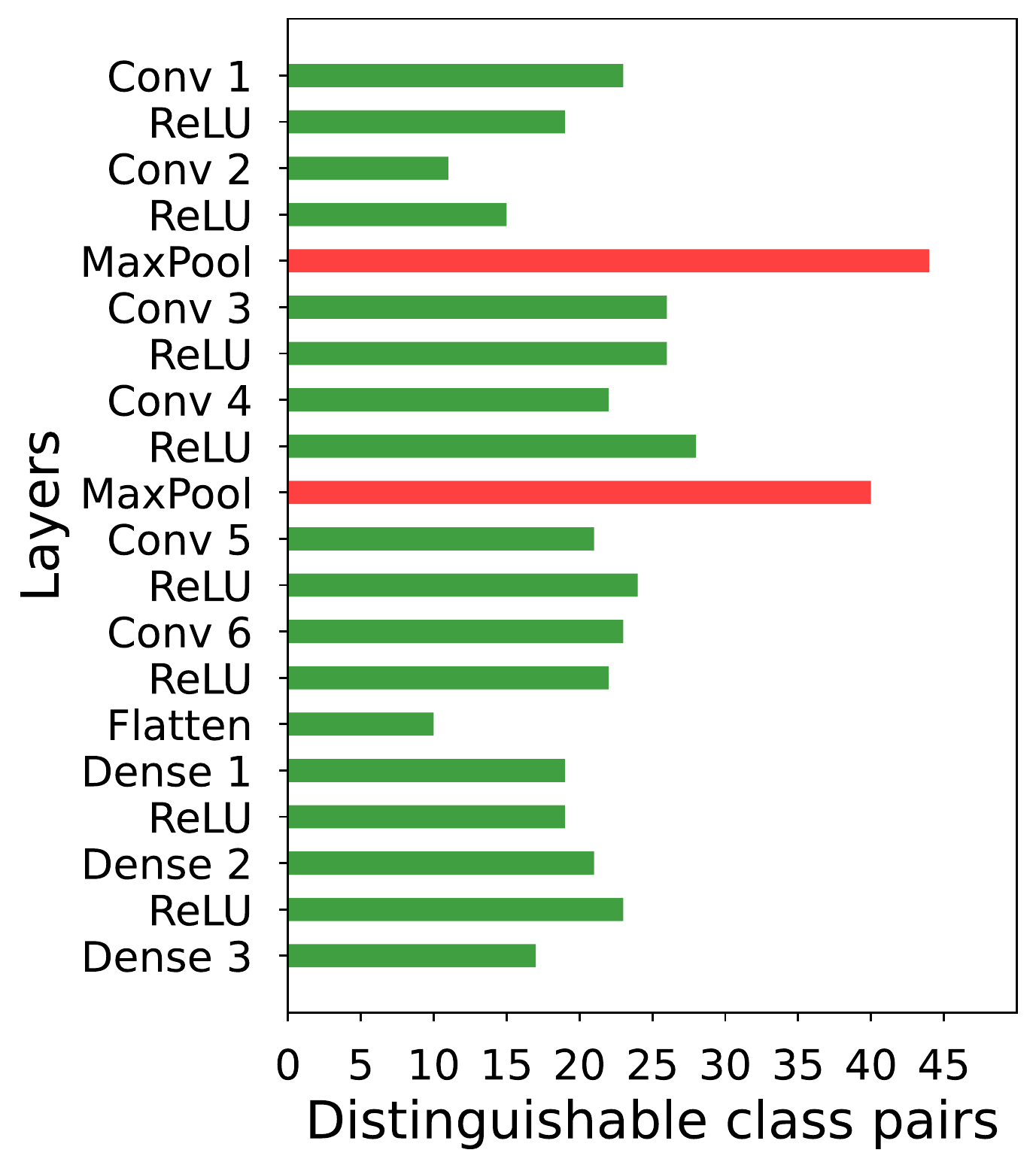}}
\caption[Number of class pairs (out of 45) distinguishable in each layer of Custom CNN trained with differential privacy(CIFAR10)]{Number of class pairs (out of 45) distinguishable in each layer of Custom CNN with Max Pooling trained with differential privacy(CIFAR10)}
\label{fig_layers_DP}
\end{figure}

To further verify that the MaxPool function is the source of leakage for differential-private CNNs as well, we do a layer-wise analysis. Following the experiment steps similar to Section \ref{section:ATM_LPTA}, the results are shown in Fig.~\ref{fig_layers_DP}. We have a total of twenty layers, which comprise of convolution, pooling, activation and fully-connected layers. On average, the six convolution layers are able to distinguish $21$ class-pairs, the eight ReLU activation distinguish $22$ class-pairs and the two dense layers distinguish $19$ class pairs, which are all less than fifty percent. On the other hand, the two Maxpool layers distinguish $44 (97.7\%)$ and $40 (88.8\%)$ class-pairs respectively. Hence, this again confirms the fact that Maxpool function is causing the timing difference, for differential-private CNNs as well. In the following discussion, we go one step further, launching the MLP class-label attack on the timing dataset of differential private custom CNN model.

\subsubsection{Attack on Differential Privacy Dataset}
\label{section:Attack_CNN_DP}
\begin{figure}[!t]
% \centerline{\includegraphics[width=0.4\textwidth]{CIFAR10_100_DP_red.png}}
\centerline{\includegraphics[width=0.6\linewidth]{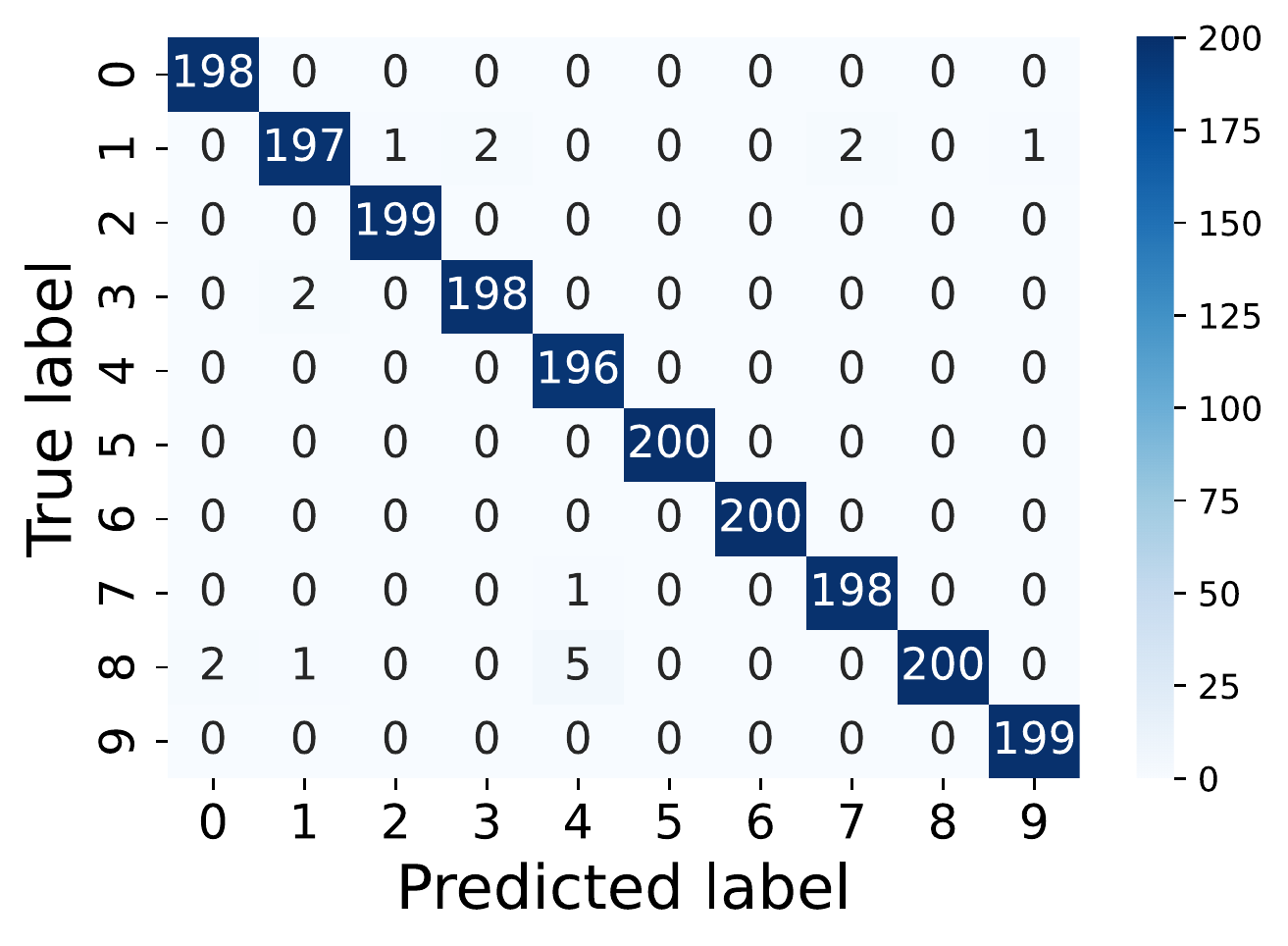}}
\caption{Confusion Matrix for timing test data on the MLP model with timing values from custom CNN trained with differential privacy}
\label{fig:cm_MLP_DP}
\end{figure}

Till now, we have verified the constant presence of PyTorch's vulnerability in differential privacy enabled CNN model from our timing analysis experiments. Our last step in this verification process is to check the effectiveness of our label classifier attack (Section \ref{section:MLP}). We follow the same steps to create the dataset and train our MLP model as we did in Section \ref{section:MLP}. Our attack gives a high accuracy score of 99.2\%, and we can now be sure that the noise added by Opacus does not have any impact on timing leakage. The test data we use has a total of $1000$ data points, $200$ for each class-label. Fig.~\ref{fig:cm_MLP_DP} shows the confusion matrix for the test data classification, which gives information about the \emph{true label} of any class and also the \emph{predicted label} of that class by the classifier. From the figure, we can see that for all classes we get more than 98\% class-label classification accuracy. In the next section, we finally see how to leak information about the training set by bypassing differential privacy.

% The figure shows that classes  5, 6 and 7 give 100\% accuracy classfication results and remaining classes have less than 2\% misclassfication rate.

% In section \ref{section:MLP} we saw that the MLP classifier is able to cause a privacy breach by being able to correctly classify corresponding class labels of different inputs, we went ahead to check whether this breach can be handled by implementing differential privacy in our original model. We use the Opacus library to train the differentially private custom CNN model and create the inference time dataset for the MLP model. We get a  high test accuracy of 99.2\% which indicates that even after adding noise to the model using Opacus the vulnerability still remains. Figure \ref{fig:cm_MLP}
% (b) shows the confusion matrix for the test data classification.

\subsection{Threat Model}

\begin{figure}[!t]
% \centerline{\includegraphics[width=0.45\textwidth]{ThreatModel_multiple_clients.png}}
\centerline{\includegraphics[width=0.4\textwidth]{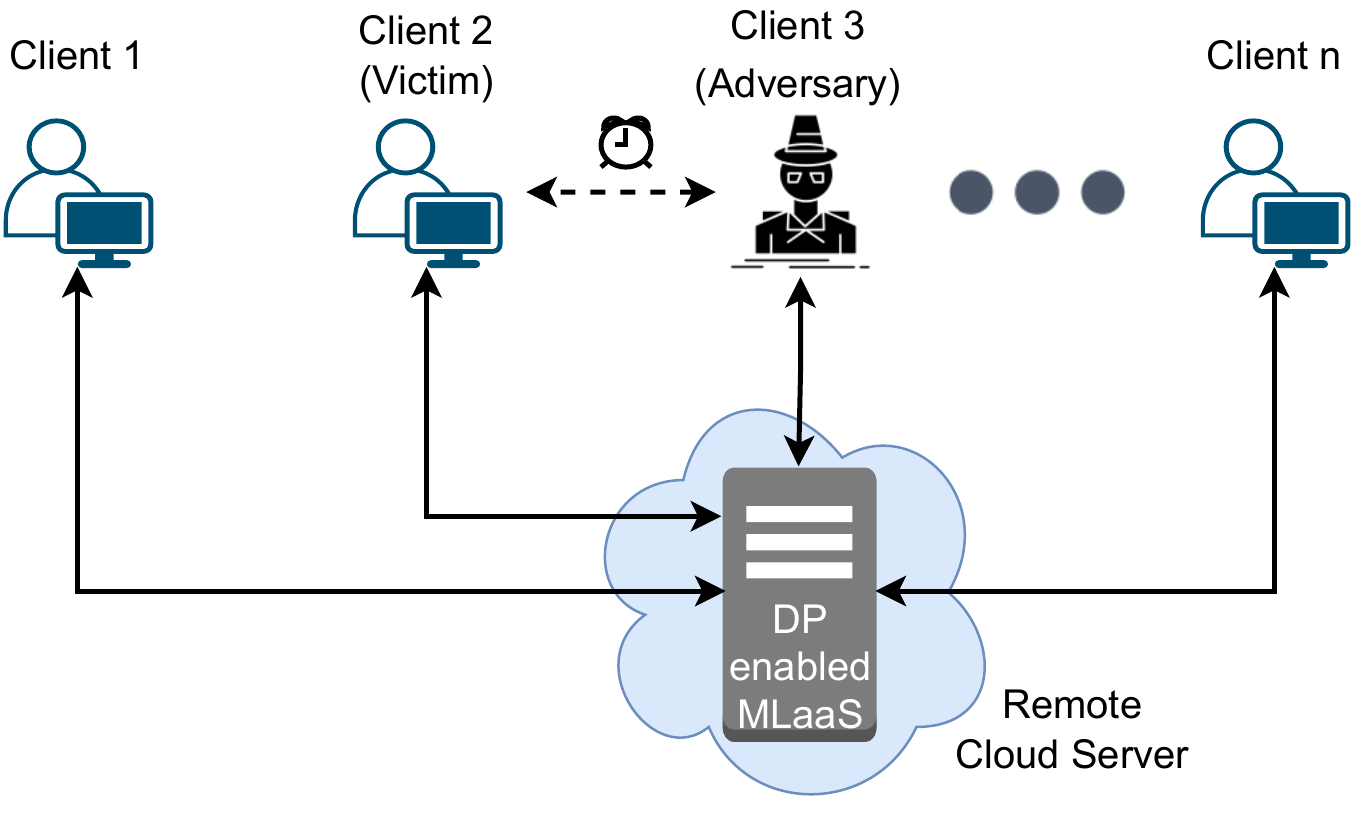}}
\caption{Threat Model for Privacy Violation in DP Models}
\label{threat_model_dp}
\end{figure}

The threat model is typically a real-life scenario of multiple clients accessing a cloud server that provides MLaaS where one amongst the many clients is considered to be adversarial, but this time the DL model on the server is trained with differential privacy which protects training data information leakage (Refer to Fig.~\ref{threat_model_dp}). Additionally, the private training data for the model is fed by the victim client accessing MLaaS, whereas the adversary tries to gain information about the training data.

\subsubsection*{Adversary Capabilities and Objective}

The adversary co-resides with the victim client on the MLaaS server and has hard-label black-box access to the DL model with user-level privilege. The adversary wishes to demonstrate the violation of Differential Privacy (DP) by ascertaining whether a dataset $Q$ has been used in the training process. The adversary interacts with a victim client to observe the timing required for training with its own set, say $T$, and build the target model, say $M_1$. Subsequently, the adversary obtains timing data of $Q$ by feeding it into the model $M_1$. The adversary then adaptively updates $M_1$ by providing the inputs consisting of both $T$ and $Q$ and builds the model $M_2$. The adversary now obtains the timing data for inferring $Q$ by model $M_2$. The attacker's objective is to determine whether $Q$ has been used in the original training data $T$. It may be emphasized that by the definition of DP, any statistics (in our case timing) gathered by simulating with differential data (in our case $T$, and $T$ augmented with $Q$), should not be distinguishable. Violation of this indicates a breach of the DP guarantees.

% \subsubsection{Practical Attack Scenario}
% Several healthcare organizations use cloud services to deploy their deep learning models, for various purposes \cite{mejia_2019}. The models mainly use datasets containing information from patient health records, with patients' identities kept anonymous for privacy reasons. We take the case where multiple clients are accessing one such deep learning model and one of the client is from a competitor company who is an adversary or spy. The adversary will also have similar dataset of patients used by his company for training DL models of their own. Now, if the adversary wishes to steal information about the training set data used by the victim healthcare company, he can launch a membership inference attack using his dataset and information from the timing channel of the victim-server communication. In this paper, we have shown how an adversary can launch a profiling attack using timing side-channels, to infer the class-labels. We also show that with the help of the same profiling attack model the adversary can know whether inputs fed by the victim belong to the training dataset or not.

\subsection{Membership Inference Attack by Bypassing Differential Privacy}
\label{section:Bypass_DP}
\begin{figure}[!t]
\centerline{\includegraphics[width=0.8\linewidth]{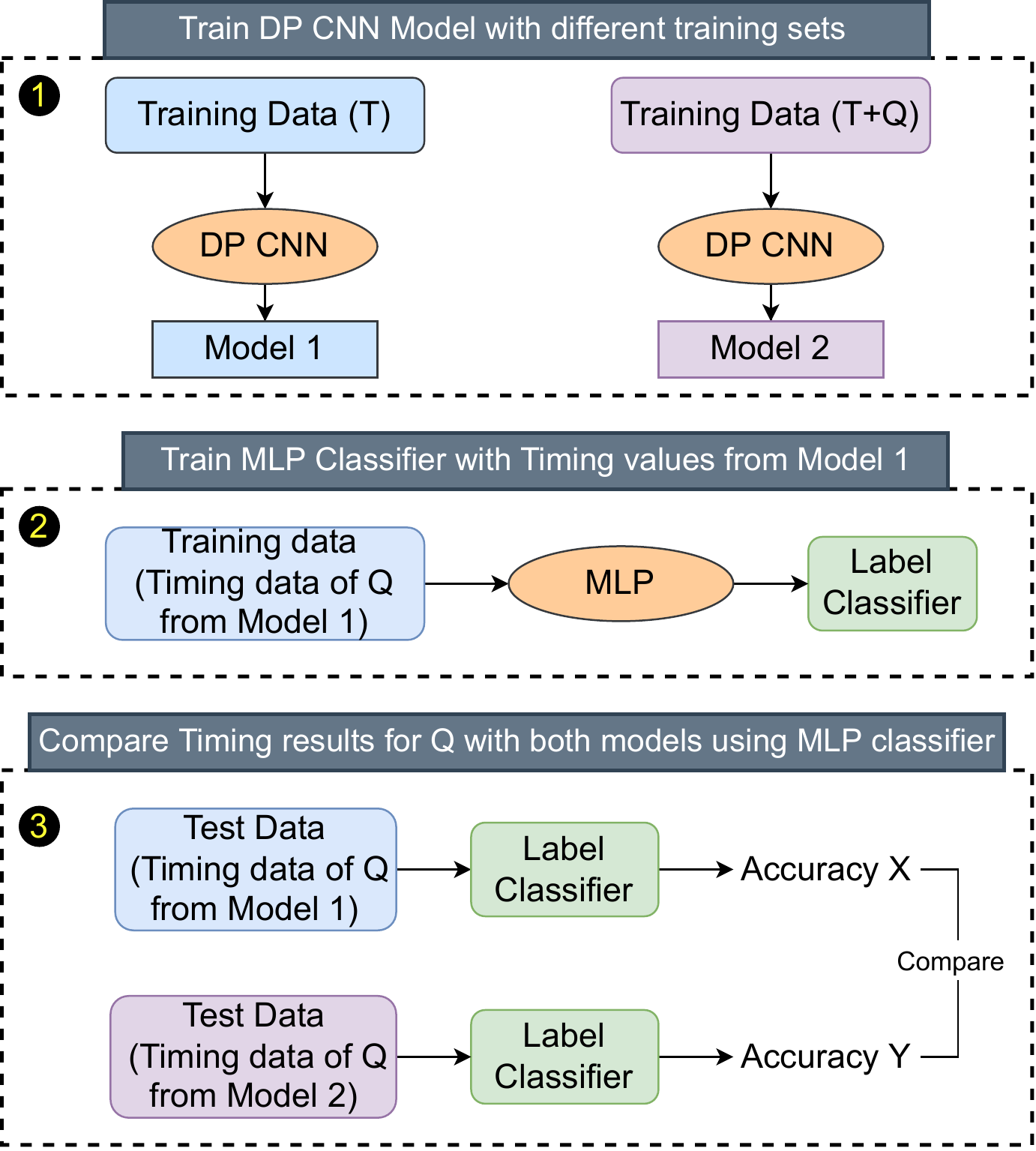}}
\caption{Bypassing Differential Privacy}
\label{MLP_ByPass}
\end{figure}

From previous experiments, it is established that the PyTorch vulnerability persists even after training DL models with differential privacy. In this section, we exploit this vulnerability for the purpose of bypassing differential privacy to mount a Membership Inference Attack (MIA). As illustrated in Fig.~\ref{MLP_ByPass}, we have divided the complete process into three parts: \emph{(i) Train DP CNN Model with different training sets: } The adversary takes a set $T$ of input images and sends it to the victim client for training the MLaaS model using this dataset, and call it as \emph{Model 1}. Next, the adversary takes a set $Q$ of input images and gives it to \emph{Model 1} for classification and the inference time to create the training and test datasets for the MLP classifier. Let us call the test set as \emph{S1}. The adversary now sends the data $Q$ to the victim for it to adaptively update the MLaaS model by training it with additional data, and calling it \emph{Model 2}. Once again, the adversary collects the inference time of set $Q$ using \emph{Model 2} to create a new test set for the MLP classifier and call it \emph{S2}. In an ideal scenario when there is no timing-leakage, the inference time for images of set $Q$ should be similar with both \emph{Model 1} and \emph{Model 2}. \emph{(ii) Train MLP Classifier with Timing values from Model 1:} Now, we have one training set created with inference time values of $Q$ from \emph{Model 1} and two test sets created using timing values of $Q$ from \emph{Model 1} and \emph{Model 2}. The adversary trains the MLP classifier using the training set, let us call it the \emph{Label Classifier}. 
\emph{(iii) Compare Timing results for Q with both models using MLP classifier: } Next, the adversary feeds both \emph{S1} and \emph{S2} to the \emph{Label Classifier} separately and compares their classification accuracy. For our experiment, we've taken $1000$ images (100 images of 10 class labels) in set $Q$ and we achieve the accuracy of 99.25\% and 82.32\% for \emph{S1} and \emph{S2}. We observe that the \emph{S2} shows a reduced accuracy, indicating towards the fact that the inference times for set $Q$ with \emph{Model 1}  and \emph{Model 2} differ from each other. Next, we do a similar analysis for the case when the test data partially overlaps with the training data.

\par
\begin{figure}[!t]
\centerline{\includegraphics[width=0.6\linewidth]{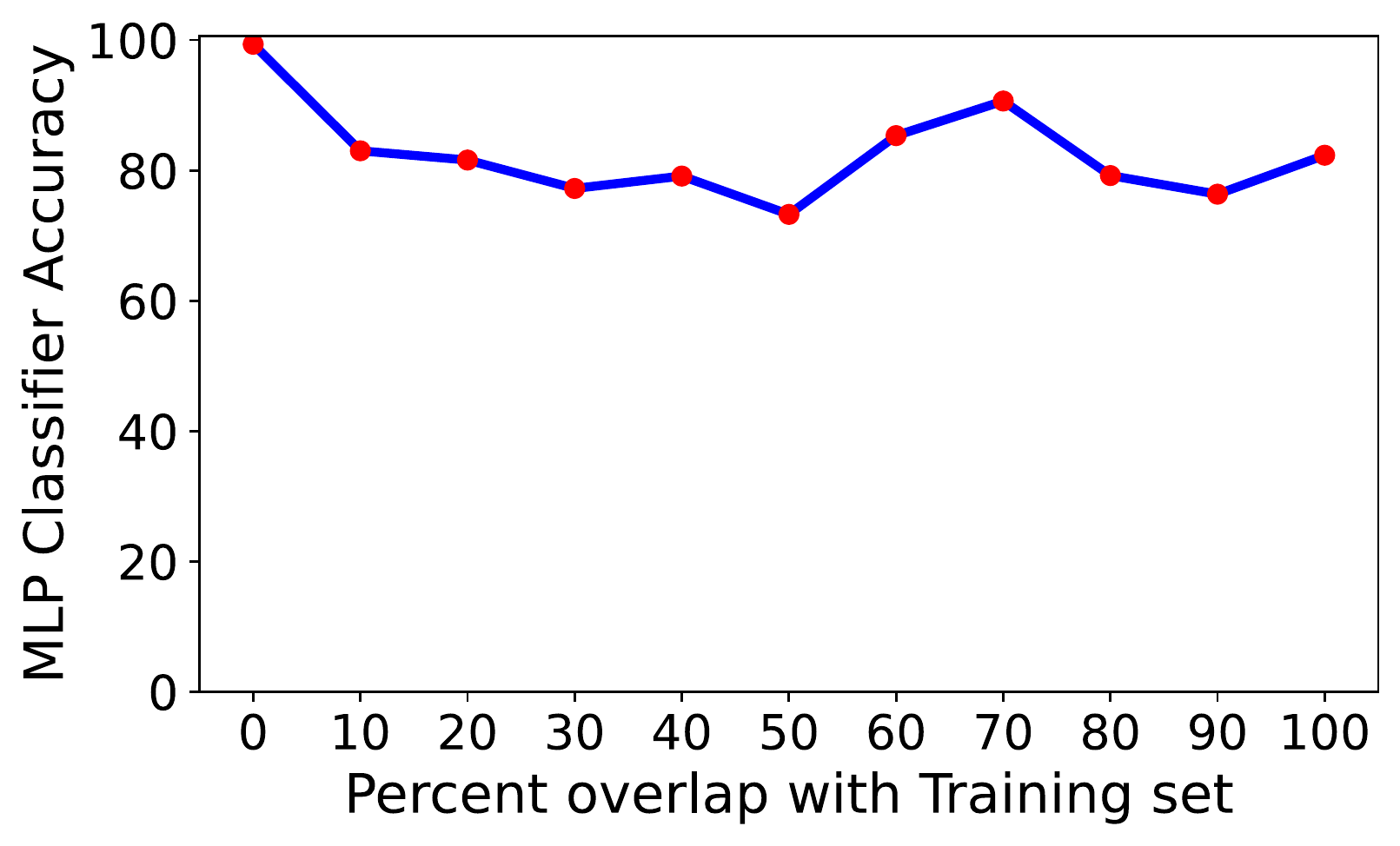}}
\caption{Accuracy of MLP Class-label classifier (trained using dataset with no overlap) over increasing ratios of training and test data overlap.}
\label{mlp_overlap_acc}~\vspace*{-0.3cm}
\end{figure}

 \textbf{Analysis with partial overlap of test and training data: }
In the previous experiment, we trained  \emph{Model 2} by adding the whole $Q$ image set and observed a drop in accuracy of the MLP classifier which was trained with timing data from \emph{Model 1}. Based on these results, we try to verify if this also happens when only a subset of images from set $Q$ are added to the training dataset of \emph{Model 2}. For our experiment, we start if $10\%$ overlap of set $Q$ with the training set and go on till $90\%$ by increasing $10\%$ overlap each time. The results are shown in Fig.~\ref{mlp_overlap_acc}. We don't see any increasing or decreasing trend in the plot, but the accuracy has dropped for by a minimum of 10\% for all overlap ratios. From this, we can infer that even a slight amount of overlap of the test set with training set will reduce the accuracy of the classifier and that can be exploited by the adversary to launch a MIA.

\par In general, we can confirm from all the results that the PyTorch vulnerability persists even after applying differential privacy to the models and can also be exploited to bypass differential privacy to leak information about the training dataset. In spite of differential privacy, Membership inference attack is also feasible with this vulnerability, hence we need a countermeasure to mitigate it. The best approach will be to mitigate it at the root level, by rectifying the Maxpool function's implementation, which we discuss in the next section.

\section{Proposed Countermeasure against Class-Leakage}
\label{section:countermeasure}

In the previous section, we explored in depth the timing vulnerabilities in PyTorch and developed attack methodologies to exploit them in a realistic framework. In this section, the vulnerability caused by the Maxpool function is thwarted by proposing an update to the existing implementation as a countermeasure. The idea is to patch PyTorch's \texttt{maxpool2d()} CPU function. PyTorch has multiple implementations of Maxpool functions (including \texttt{maxpool2d()}) for multiple types of input, devices, and applications, and this countermeasure could easily be implemented at all places. This section begins by explaining the implementation of the proposed countermeasure in PyTorch, and later results of timing analysis and MLP attack, run on mitigated PyTorch library are discussed.

\subsection{Countermeasure Implementation}
\begin{figure}[!ht]
\centerline{\includegraphics[width=0.35\textwidth]{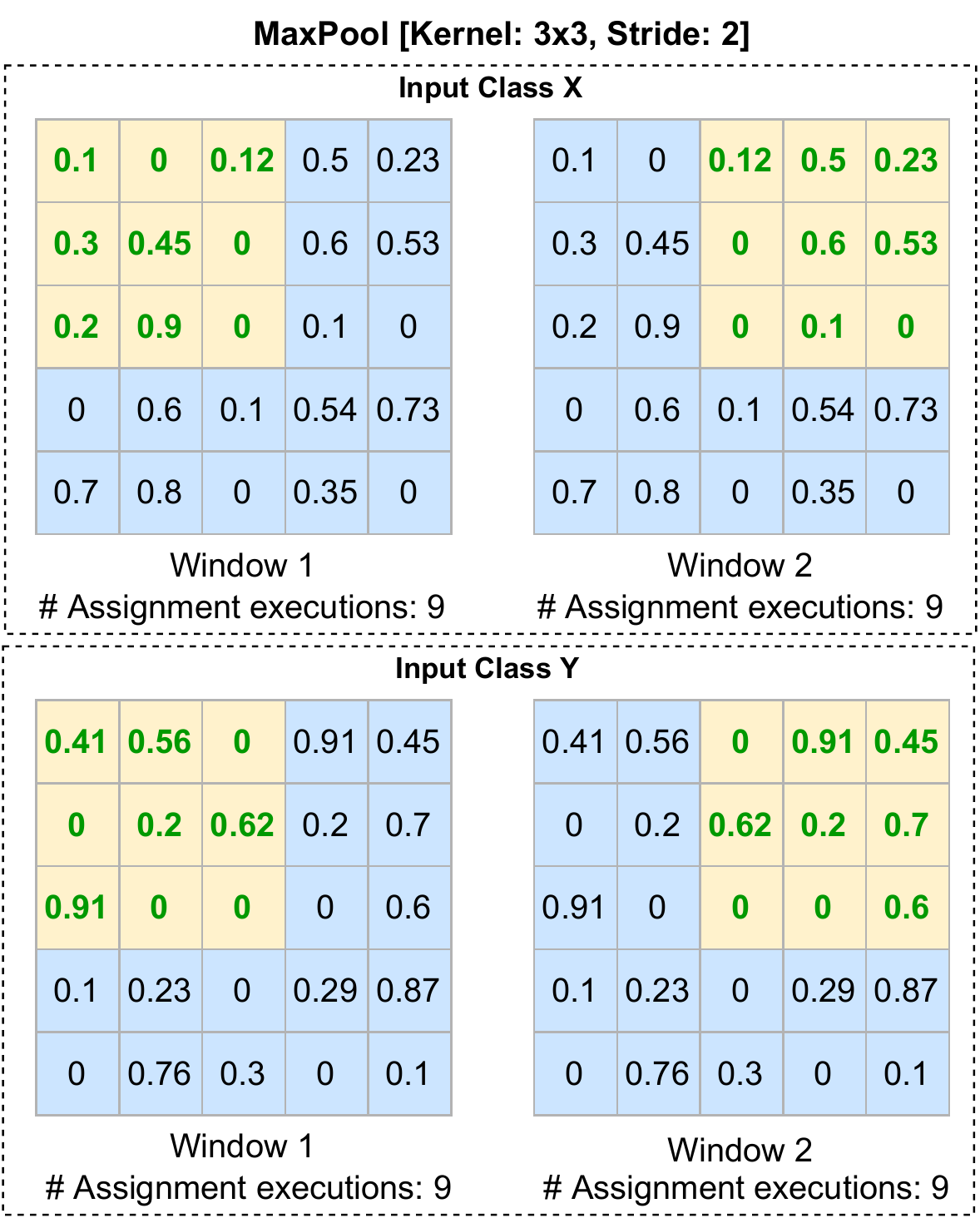}}
\caption{An example of constant number of assignment operations for two different inputs during Maxpool operation after implementing countermeasure}
\label{if_stat_eg_counter}
\end{figure}

It was shown in section \ref{section:pytorch_vul}  that Maxpool's implementation vulnerability is caused by the difference in the number of executions of assignment statements inside the `if' condition. To mitigate this, we replicated the `if' statement's functionality by adding a temporary swap location replacing the code snippet in Listing \ref{code_snippet} with Listing \ref{code_snippet_fix}. In the fixed code, an additional temporary array $tmp\_arr$ is introduced  which is declared above the outermost $for$ loop inside the $cpu\_max\_pool()$ function.  In Fig.~\ref{if_stat_eg_counter} we show that after implementing the countermeasure, assignment statement will be executed for every element of a window, hence the overall number of assignment operations inside the maxpool layer will be constant for all inputs.  The yellow window represents the current window on which the Maxpool operation is being performed, and the bold green numbers indicate the positions at which the assignment operations are executed. For this example, we see that assignment operation is run for all windows hence constant time for all inputs. With the updated implementation, the input data dependency is removed completely, and therefore we claim to get uniform inference times for all classes, which are illustrated in the next section.

\begin{lstlisting}[language=Python, caption=Proposed constant time updated code to PyTorch Maxpool code structure, label=code_snippet_fix]
tmp_arr[0] = val;
tmp_arr[1] =  maxval;
maxval = tmp_arr[(val < maxval)*1];
tmp_arr[0] = index;
tmp_arr[1] = maxindex;
maxindex = tmp_arr[(val < maxval)*1];
\end{lstlisting}

\subsection{Mitigation of the PyTorch Vulnerability}
%We've introduced the countermeasure to mitigate the timing leakage and now 
In this section we explore the efficiency of the proposed countermeasure by repeating all the timing analysis experiments done in Section \ref{section:ATM}, and the experiments were repeated using the mitigation patched PyTorch library.
% In the implemented countermeasure, all the statements will be executed for each element of the pooling window to calculate the maximum value. Hence, the maxpool function execution time will be now similar for all inputs. Consequently, the overall inference time for all inputs will also become similar.
\subsubsection{Analysing Overall Inference Time:}
\begin{figure}[!t]
% \centerline{\includegraphics[width=0.35\textwidth]{Countermeasure_before_after.png}}
\centerline{\includegraphics[width=0.5\linewidth]{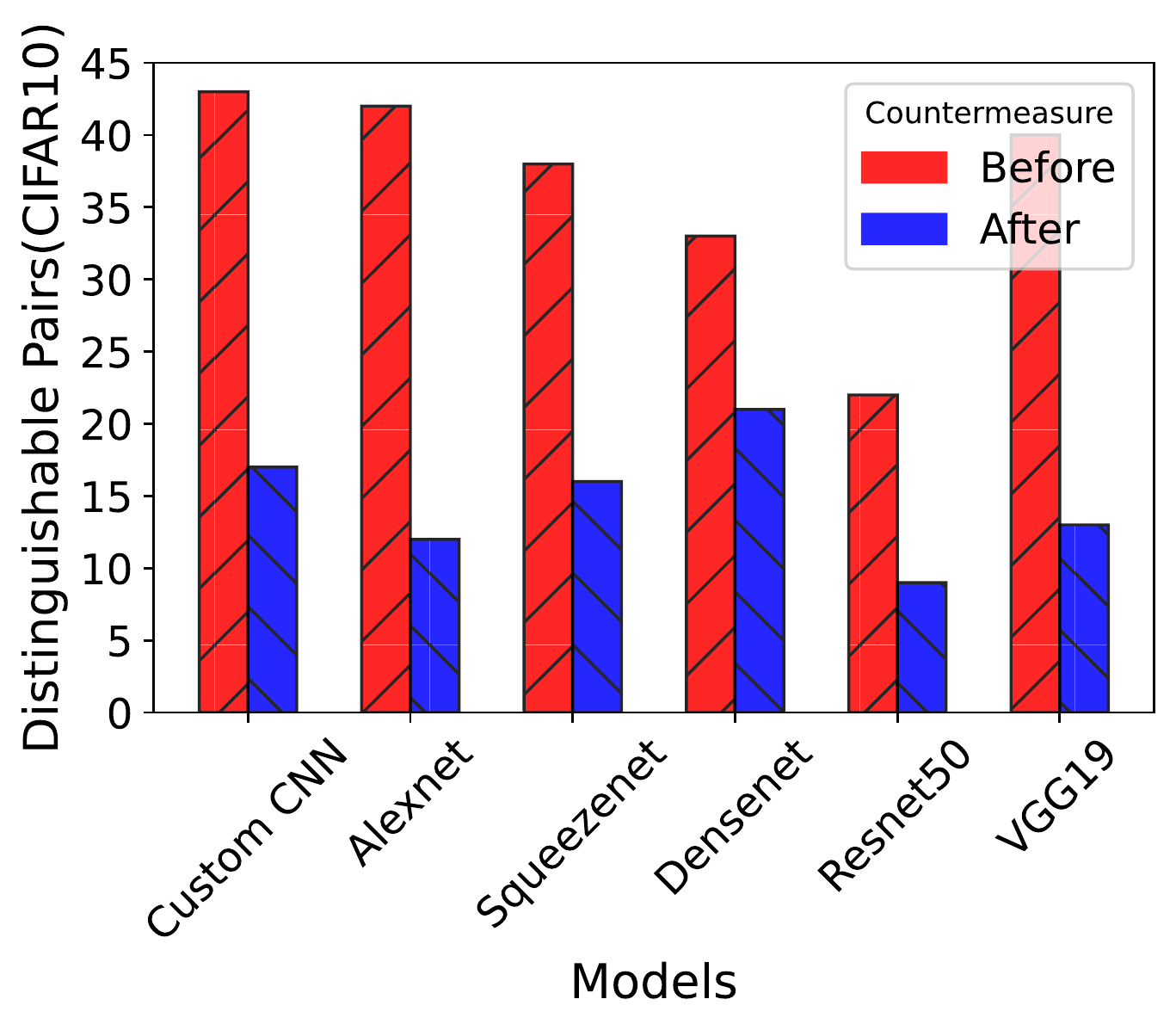}}
\caption[Number of class pairs (out of 45) distinguishable by all CNN models before and after implementing the counter measure (CIFAR10)]{Number of class pairs (out of 45) distinguishable by all CNN models before and after implementing the counter measure (CIFAR10)}
\label{before_after}
\end{figure}

%We do overall inference function  
Fig.~\ref{before_after} shows timing analysis similar as discussed in Section \ref{section:ATM_PTA} for all six models (Custom CNN, Alexnet, Resnet50, Densenet, Squeezenet and VGG19) implemented on countermeasure enabled PyTorch library. The figure shows that for all models, unlike the vulnerable implementation, the number of distinguishable pairs for the patched library has reduced to less than 50\% of the total pairs after the countermeasure implementation. From this observation, we apparently make sure that the countermeasure implementation works as it claims to be. Next, we dig deep with layer-wise analysis, to see the timing behavior of the countermeasure on the exact source of leakage.

\subsubsection{Analysing Layer-wise Inference Time}
\begin{figure}[!t]
\centerline{\includegraphics[width=0.5\linewidth]{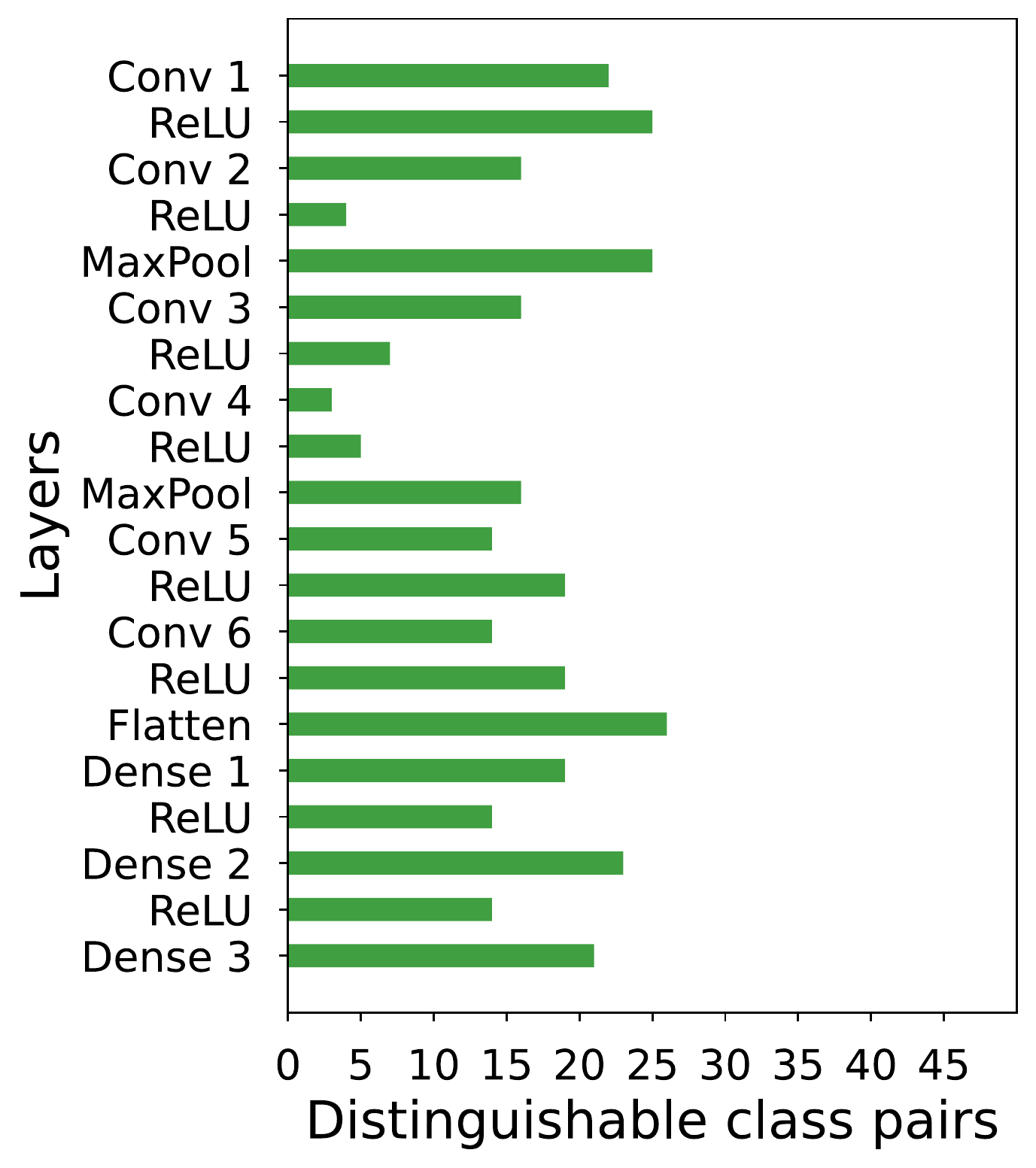}}
\caption[Number of class pairs (out of 45) distinguishable in each layer of Custom CNN (CIFAR10)]{Number of class pairs (out of 45) distinguishable in each layer of Custom CNN with Max Pooling after implementing the countermeasure (CIFAR10)}
\label{fig_layers_fix}
\end{figure}

Having a working countermeasure we verify its effectiveness using the layer-wise analysis by following the experiment steps from Section \ref{section:ATM_LPTA}. In Fig.~\ref{fig_layers_fix} we observe that the number of distinguishable pairs in the max pool layers has decreased to 25 and 16 from 45 and 44 in Fig.~\ref{fig:fig_layers}(a). In the upcoming section, we verify the countermeasure's compatibility with differential-private networks.

\subsection{Mitigation of Class-Leakage in Differential Privacy}
In Section \ref{section:Bypass_DP}, we demonstrated how differential privacy can be bypassed by exploiting Python's vulnerability, hence it becomes important to verify whether the countermeasure also works for the PyTorch models which are trained with differential privacy using the Opacus library.
\par
\begin{figure}[!t]
% \centerline{\includegraphics[width=0.35\textwidth]{Countermeasure_before_after_DP.png}}
\centerline{\includegraphics[width=0.5\linewidth]{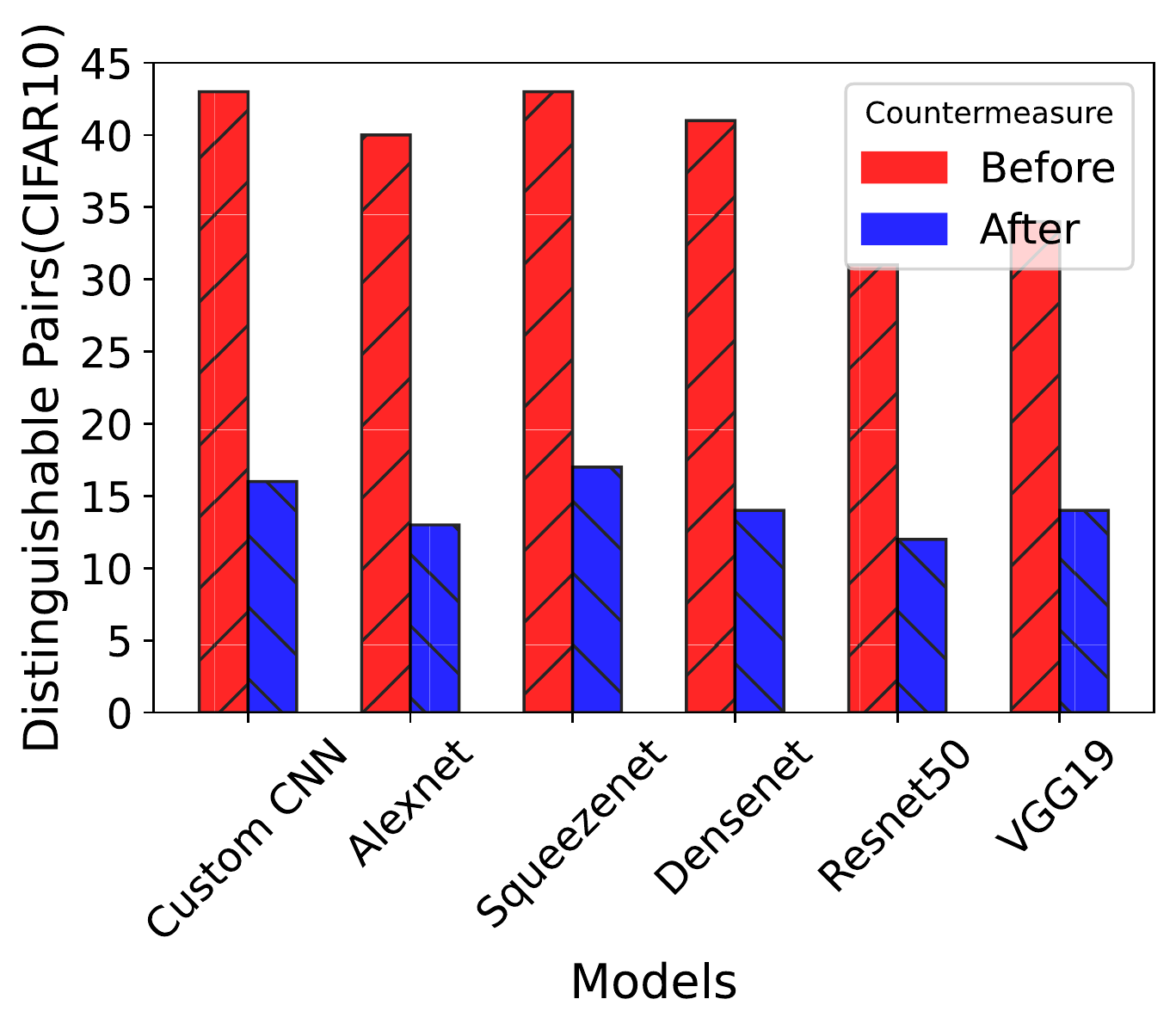}}
\caption[Number of class pairs (out of 45) distinguishable by all CNN models trained with differential privacy before and after implementing the counter measure (CIFAR10)]{Number of class pairs (out of 45) distinguishable by all CNN models trained with differential privacy before and after implementing the counter measure (CIFAR10)}
\label{before_after_DP}
\vspace{-0.4cm}
\end{figure}

 \emph{Analysing Overall Inference Time}
Once again we start by analysing the overall inference timing for all six differential-private models (Custom CNN, Alexnet, Resnet50, Densenet, Squeezenet, and VGG19).  In Fig.~\ref{before_after_DP}, the number of distinguishable pairs once again reduces to less than 50\% of the total pairs for all the models, hence confirming the effectiveness of our countermeasure for differential-private networks as well. In the following section, we explore the MLP profiling attack with countermeasure enabled in PyTorch.

\subsection{Attack with Countermeasure}
In Section~\ref{section:Attack_CNN} and Section~\ref{section:Attack_CNN_DP} we saw that MLP class-label classifier gave high accuracy for custom CNN model trained with and without differential privacy. Now, in this section, we again launch the attack with implemented countermeasure.
\subsubsection{Attack on Custom CNN Dataset with Countermeasure}
To confirm the effectiveness of our countermeasure, a new dataset for the MLP is created with the mitigated custom CNN model using the steps we discussed in Section \ref{section:MLP}. The MLP is trained with the dataset and then the classes of test data is inferred. Results show a tremendous drop in the accuracy from 99.35\% to 12\%, hence proving the viability of our countermeasure.

\subsubsection{Attack on Differential Privacy Dataset with Countermeasure}
\begin{figure}[!t]
% \centerline{\includegraphics[width=0.45\textwidth]{MLP_accuracy.png}}
\centerline{\includegraphics[width=0.4\linewidth]{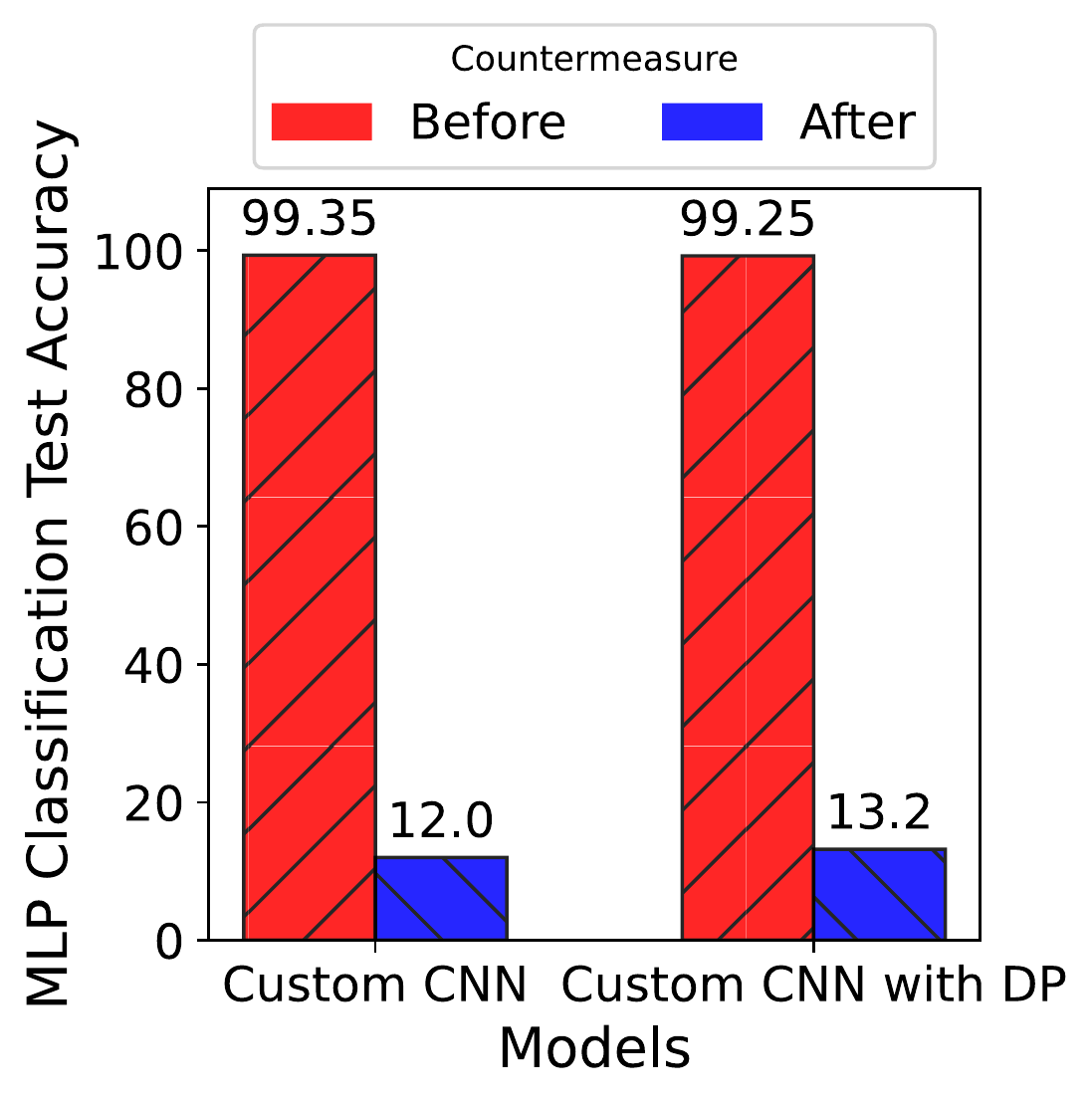}}
\caption[Accuracy of MLP model using inference time dataset of Custom CNN model]{Accuracy of MLP model using inference time dataset of Custom CNN model}
\label{mlp_acc}
\vspace{-0.4cm}
\end{figure}

Next follows the efficacy of our countermeasure on the 
dataset created using differential private custom CNN. Once again we see a drop in accuracy from 99.25\% to 13.2\%  which indicate towards random classification, meaning the model is no more able to classify among the different class label inputs. This shows that our countermeasure works perfectly in all scenarios. Fig.~\ref{mlp_acc} shows the comparative results for attack on both normal as well as differential private datasets.

% ----------------

\section{Vulnerability Disclosure}
\label{section:fb_ack}
We reported the observed data-dependent timing side-channel leakage due to improper non-constant time implementation of Max Pooling operation in PyTorch (i.e., the discussion presented in Section~\ref{section:vulnerability}) to Meta (Facebook) AI research (developer of the PyTorch library) on 16 January 2022. We received an acknowledgment for reporting a valid issue on 14 February 2022. Quoting their exact words:
\begin{quoting}
\textit{We have discussed the issue at length and concluded that, whilst you reported a valid issue which the team may make changes based on, unfortunately your report falls below the bar for a monetary reward.}
\end{quoting}
We further reported the practical attack on user privacy exploiting the data-dependent timing side-channel leakage (i.e., the discussion presented in Section~\ref{section:MLP_Attack}), the vulnerability of differential private deep learning models against membership inference attacks exploiting the same timing-channel (i.e., the discussion presented in Section~\ref{section:DP_CL_CNN}), and the countermeasure to alleviate the data-dependent timing side-channel (i.e., the discussion presented in Section~\ref{section:countermeasure}).

\section{Discussion and Conclusion}
\label{section:conclusion}
PyTorch's user base has exponentially increased since its release in 2016. Hence, the timing leakage vulnerability observed in the PyTorch library is a serious privacy concern for the parties using it to work with highly confidential data, and it should not be ignored. Major technology giants have built their deep learning models on PyTorch including Microsoft, Tesla, Uber, Airbnb, and Facebook itself. Tesla has built its Autopilot~\cite{DBLP:conf/smc/DikmenB17} system on PyTorch. The adversary could actually get hold of the timing leakage to get the classification result of the models running to predict the car's next move. The predictions can be put together to get the complete route taken by the car. This is a potential scenario of serious privacy threat for the car's owner as well as Tesla. Further, this leakage is not only possible with timing side-channel but also realisable with hardware performance counter events such as branching instruction leaking similar information about the underlying implementation vulnerability. Another possible threat scenario is for healthcare organizations that use cloud services to deploy their deep learning models, for various purposes~\cite{mejia_2019}. The models mainly use datasets containing information from patient health records, with patients' identities kept anonymous for privacy reasons. Using the MIA attack discussed in Section~\ref{section:Bypass_DP} an adversary from some rival company can steal information about the training set data used by the victim healthcare company, by bypassing differential-private models and breaching patients' privacy.

% We take the case where multiple clients are accessing one such deep learning model and one of the client is from a competitor company who is an adversary or spy. The adversary will also have similar dataset of patients used by his company for training DL models of their own. Now, if the adversary wishes to steal information about the training set data used by the victim healthcare company, he can launch a membership inference attack using his dataset and information from the timing channel of the victim-server communication. In this paper, we have shown how an adversary can launch a profiling attack using timing side-channels, to infer the class-labels. We also show that with the help of the same profiling attack model the adversary can know whether inputs fed by the victim belong to the training dataset or not.

\par In this paper we bring forward for the first time, a potential privacy threat found in the PyTorch library caused by timing side-channel leakage, which an adversary can exploit to get the input class of the data being fed to a neural network. The source of leakage was diagnosed to be the Maxpool layer of the network. The vulnerability caused by this leakage can be exploited to classify class labels of the inputs by training a MLP classifier with statistically processed inference time dataset. This was further utilized using the MLP classifier to bypass differential privacy to identify whether a set of inputs to the model belongs to CNN model's training dataset. Finally, we propose an inexpensive yet effective implementation as a countermeasure to thwart such timing vulnerability.

% \bibitem{b1} G. Eason, B. Noble, and I. N. Sneddon, ``On certain integrals of Lipschitz-Hankel type involving products of Bessel functions,'' Phil. Trans. Roy. Soc. London, vol. A247, pp. 529--551, April 1955.
% \bibitem{b2} J. Clerk Maxwell, A Treatise on Electricity and Magnetism, 3rd ed., vol. 2. Oxford: Clarendon, 1892, pp.68--73.
% \bibitem{b3} I. S. Jacobs and C. P. Bean, ``Fine particles, thin films and exchange anisotropy,'' in Magnetism, vol. III, G. T. Rado and H. Suhl, Eds. New York: Academic, 1963, pp. 271--350.
% \bibitem{b4} K. Elissa, ``Title of paper if known,'' unpublished.
% \bibitem{b5} R. Nicole, ``Title of paper with only first word capitalized,'' J. Name Stand. Abbrev., in press.
% \bibitem{b6} Y. Yorozu, M. Hirano, K. Oka, and Y. Tagawa, ``Electron spectroscopy studies on magneto-optical media and plastic substrate interface,'' IEEE Transl. J. Magn. Japan, vol. 2, pp. 740--741, August 1987 [Digests 9th Annual Conf. Magnetics Japan, p. 301, 1982].
% \bibitem{b7} M. Young, The Technical Writer's Handbook. Mill Valley, CA: University Science, 1989.
% \end{thebibliography}
% \vspace{12pt}
% \color{red}
% IEEE conference templates contain guidance text for composing and formatting conference papers. Please ensure that all template text is removed from your conference paper prior to submission to the conference. Failure to remove the template text from your paper may result in your paper not being published.

\bibliographystyle{IEEEtran}
\bibliography{sample}
\end{document}